\documentclass[11pt, a4paper]{article}
\pdfoutput=1
\usepackage[utf8]{inputenc}
\usepackage[T1]{fontenc}
\usepackage{lmodern, textcomp}
\usepackage[american]{babel}
\usepackage{csquotes}
\usepackage{bbold}
\usepackage{multicol,caption}

\usepackage{mathtools}

\usepackage{eurosym}
\usepackage
[nottoc]{tocbibind}
\usepackage{authblk}
\usepackage{fancyhdr}
\usepackage{xspace}
\usepackage{bm}

\usepackage{graphicx}
\usepackage{subcaption}
\usepackage{float}
\usepackage{makecell}
\usepackage{multirow}
\usepackage{multicol} 
\usepackage{marginnote}
\usepackage[multiple]{footmisc}

\usepackage[usenames, dvipsnames, svgnames, table]{xcolor}

\usepackage[backend=bibtex8, citestyle=numeric-comp, maxbibnames=99,
			sorting=none, sortcase=false, sortcites=true, giveninits=true]{biblatex}

\usepackage{amsmath, amsfonts, amssymb}
\usepackage{mathtools}
\usepackage{mathrsfs}
\usepackage{cancel}
\usepackage{braket}
\usepackage{tensor}
\usepackage{slashed}
\usepackage{siunitx}
\usepackage{listings}

\usepackage{stmaryrd}
\usepackage{caption}
\usepackage{relsize,exscale}
\usepackage{array}
\usepackage[toc,page]{appendix}

\usepackage{enumerate}

\DeclareSymbolFont{largesymbols}{OMX}{cmex}{m}{n}

\setcellgapes{1pt}
\makegapedcells
\newcolumntype{R}[1]{>{\raggedleft\arraybackslash }b{#1}}
\newcolumntype{L}[1]{>{\raggedright\arraybackslash }b{#1}}
\newcolumntype{C}[1]{>{\centering\arraybackslash }b{#1}}

\newcommand{\Tr}{\mathrm{Tr}}

\newtheorem{definition}{Definition}

\newtheorem{proposition}{Proposition}
\newtheorem{remark}{Remark}
\newtheorem{claim}{Claim}

\newtheorem{lemma}{Lemma}

\newcommand{\beq}{\begin{equation}}
\newcommand{\eeq}{\end{equation}}
\newcommand{\bea}{\begin{eqnarray}}
\newcommand{\eea}{\end{eqnarray}}
\definecolor{mygray}{gray}{0.3}

\newcommand{\bes}{\begin{eqnarray}}
\newcommand{\ees}{\end{eqnarray}}

\newcommand\restr[2]{{
  \left.\kern-\nulldelimiterspace 
  #1 
  \vphantom{\big|} 
  \right|_{#2} 
  }}
\def\extd{\mathrm {d}}

\newcommand{\N}{\mathrm{N}}

\newcommand{\RE}{\mathrm{Re}}
\newcommand{\IM}{\mathrm{Im}}

\usepackage{multicol}
\usepackage{amssymb}
\def\Xint#1{\mathchoice
   {\XXint\displaystyle\textstyle{#1}}%
   {\XXint\textstyle\scriptstyle{#1}}%
   {\XXint\scriptstyle\scriptscriptstyle{#1}}%
   {\XXint\scriptscriptstyle\scriptscriptstyle{#1}}%
   \!\int}
\def\XXint#1#2#3{{\setbox0=\hbox{$#1{#2#3}{\int}$}
     \vcenter{\hbox{$#2#3$}}\kern-.5\wd0}}

\def\dashint{\Xint-}

\usepackage[heightrounded, top=4cm, bottom=4cm, left=3cm, right=3cm, headheight=14pt]{geometry}

\usepackage[pdftex, breaklinks=true, linktocpage,
	colorlinks=true, urlcolor=blue, linkcolor=blue, citecolor=red
]{hyperref}

\newcommand{\email}[1]{\href{mailto:#1}{\nolinkurl{#1}}}
\newcommand{\emailfoot}[1]{\thanks{\email{#1}}}

\usepackage{marginnote}
\newcounter{draftcommentcnt}
\NewDocumentCommand{\draftcomment}{s O{red} m}{%
	\def\margnote{\IfBooleanTF{#1}{\marginnote}{\marginpar}}%
	\stepcounter{draftcommentcnt}%
	\textcolor{#2}{#3}%
	\margnote{\textcolor{#2}{$\Leftarrow$ \arabic{draftcommentcnt}}}%
}

\fancypagestyle{preprint}{
	\fancyhf{}

	\fancyhead[R]{\textsf{\preprint}}
}

\numberwithin{equation}{section}

\hypersetup{
	pdftitle={Functional renormalization group for $p=2$ matrix dynamics in the planar approximation},
}

\title{Functional renormalization group for ‘‘$p=2$'' like glassy matrices in the planar approximation\\
\bigskip
\Large{II. Ward identities method in the deep IR}}

\author[1]{Vincent Lahoche\emailfoot{vincent.lahoche@cea.fr}}
\author[1,2]{Dine Ousmane Samary\emailfoot{dine.ousmanesamary@cipma.uac.bj}}

\affil[1]{%
	Université Paris Saclay, \textsc{Cea}, \textsc{List}, Gif-sur-Yvette, F-91191, France
}
\affil[2]{%
	Faculté des Sciences et Techniques (ICMPA-UNESCO Chair)
	\protect\\
	Université d'Abomey-Calavi, 072 BP 50, Bénin
}
\begin{document}
\maketitle

\hrule
\hrule
\begin{abstract}
This paper, as a continuation of our previous investigation [arXiv:2403.07577] aims to study the glassy random matrices with quenched Wigner disorder. In this previous work,  we have constructed a renormalization group based on the effective deterministic kinetic spectrum emerging from large $\N$ limit, and we extended approximate solutions using standard vertex expansion, at the leading order of the derivative expansion. Now in the following work, by introducing the non-trivial Ward identities which come from the $(\mathcal{U}(\N))^{\times 2}$ symmetry broken of the effective kinetic action, we provide in the deep IR the explicit solution of the functional renormalization group for a model with quartic coupling by solving the Hierarchy to all orders in the local sector, which in particular imply the vanishing of the anomalous dimension. The numerical investigations confirm the first-order phase transition discovered in the vertex expansion framework, both in the active and passive schemes. Finally, we extend the discussion to hermitian matrices.

\end{abstract}

\hrule

\hrule
\newpage
\pdfbookmark[1]{\contentsname}{toc}
\tableofcontents
\pagebreak

\section{Introduction}

In our recent investigation \cite{lahoche20241}, we have considered a stochastic complex $\N\times \N$ matrix $M$, characterized by a quenched disorder, materialized by a Wigner Hermitian matrix of size $\N$. In the large $\N$ limit, we showed that the equilibrium states look like a non-conventional Euclidean and non-local field theory, with an effective kinetics spectrum given by the Wigner law. Following the strategy developed in \cite{lahoche2022generalized,lahoche2023functional}, we constructed a renormalization group based on a partial integration procedure of degrees of freedom, accordingly with the nontrivial notion of scale provided by the effective kinetics. There are three main difficulties with this effective field theory method in the point of view of the nonperturbative renormalization group techniques:
\begin{enumerate}
\item The intrinsic scale dependencies of the canonical dimension.
\item The non-local nature of the interactions.
\item The absence of an easily summable leading order sector.
\end{enumerate}
The third point is specific to the matrix case, vectors and tensors being essentially \textit{branched polymers} at the leading order \cite{guruau2017random,zinn1998vector}. Note that the restriction to the leading order is enforced because we use the Wigner law for the effective kinetics spectrum. For random matrices, the leading order sector corresponds to the planar Feynman graphs, encoding dual triangulations of surfaces with zero genus \cite{Francesco_1995}. The first point is because the momenta distribution is a power law only asymptotically. To deal with these difficulties, we essentially focused on the deep IR, where the shape of the momenta distribution matches with a three-dimensional Euclidean field theory. Because of the second point, many popular methods usually considered in the functional renormalization group literature fail \cite{Berges_2002}. In the previous work, we examined the relevance of the vertex and derivative expansion, whose relevance for nonlocal field theories has been established for tensorial field theories \cite{lahoche2016renormalization,benedetti2016functional,Carrozza_2016ccc,Carrozza_2017a,Carrozza_2017}. In this framework we considered two different scaling, that we called respectively ‘‘active'' and ‘‘passive'' schemes, and provided some evidences in favor of a first-order phase transition between a dilute phase and a condensed phase for eigenvalues of the Hermitian matrix $\chi=M^\dagger M$. 
\medskip

In this paper we are aiming to go beyond the vertex and derivative expansions, exploiting the non-trivial Ward identities arising because of the non-trivial kinetics, that break the global unitary invariance $(\mathcal{U}(N))^{\times 2}$ of the interactions. It is well known that Ward identities play a significant role in the renormalizability proofs and the construction of the renormalization group for Gauge theories \cite{Zinn-Justin:1989rgp,ZinnJustinBook2}. Ward identities have also played a significant role in the tensorial field theories context \cite{Lahoche:2018oeo,Lahoche:2018ggd}, characterized by similar non-localities as the model we consider in this paper. For these theories, the anomalous dimension receives corrections at one loop order, and the standard local vertex expansion fails to take into account the momenta dependency of the effective vertices, which, in contrast with ordinary field theories are significant. 
\medskip

The outline of this paper is the following. After a short recalling of definitions and main results obtained in the previous paper in sections \ref{sectiondef} and \ref{seceqscheme2}, we derive and investigate the Ward identities in section \ref{closing1} both for complex and hermitian matrices, in the deep IR. The main statement enforces the evidence in favor of the existence of a first-order phase transition, and we point out some opened issues in section \ref{concluding}. Some additional materials are given in Appendices \ref{App7}, \ref{App6}, and \ref{App1} to make the paper as self-consistent as possible.

\section{The model and definitions}\label{sectiondef}

Let us briefly summarize the definition of the model. We consider a complex stochastic matrix $M(t)\in \mathbb{C}^{\N\times \N}$, whose entries evolves accordingly with the stochastic process:
\begin{equation}
\frac{\extd M_{ij}}{\extd t}=-\frac{\delta \mathcal{H}}{\delta \overline{M}_{ij}(t)}+B_{ij}(t)\,,\label{eq1}
\end{equation}
where $B(t)$ is a Gaussian white noise, with zero mean and variance,
\begin{equation}
\langle \bar{B}_{i^\prime j^\prime}(t^\prime) B_{ij}(t) \rangle= T\delta_{ii^\prime}\delta_{jj^\prime}\delta(t-t^\prime)\,,
\end{equation}
and $\mathcal{H}$ reads as:
\begin{align}
\nonumber\mathcal{H}&:=\int_{-\infty}^{+\infty} \extd t\,\Tr \left( J M(t) M^\dagger(t) \right)+\Tr \left( K M^\dagger(t) M(t) \right)\\
&\quad+\sum_{p=1}^\infty\,\frac{a_p \N^{-p+1}}{(p!)^2} \,\int_{-\infty}^{+\infty} \extd t \,\Tr \, ({M}^\dagger(t)M(t))^p\,,
\end{align}
where both $J$ and $K$ are Hermitian Wigner matrices \cite{potters2020first,forrester2010log} with the same variance $\sigma^2$. In the large $\N$ limit, the spectrum of the Wigner matrices becomes deterministic, given by the well-known semi-circle law, and for any suitable test function $f$, we have the equality for $\N\to \infty$:
\begin{equation}
\frac{1}{\N}\lim_{\N \to \infty} \sum_{\mu=1}^\N\, f(\lambda_\mu)= \int_{-2\sigma}^{2\sigma} \, \extd \lambda \, \mu_W(\lambda) f(\lambda)\,,
\end{equation}
with:
\begin{equation}
\mu_W(\lambda)=\frac{\sqrt{4\sigma^2-\lambda^2}}{2\pi \sigma^2}\,.
\end{equation}
Then, at equilibrium, the large $\N$ partition function $Z_{\text{eq}}^\mathbb{C}[L,\overline{L}]$ reads setting $T=2$:
\begin{equation}
Z_{\text{eq}}^\mathbb{C}[L,\overline{L}]\underset{\N\to \infty}{\longrightarrow}\int \extd M \extd \overline{M}\, e^{- H_\infty^{\mathbb{C}}[M,\overline{M}]+ \overline{L}\cdot M+\overline{M}\cdot L}\,,\label{parteqcomplex2}
\end{equation}
where:
\begin{equation}
H_\infty^{\mathbb{C}}[M,\overline{M}]:= \sum_{\lambda,\mu} \overline{M}_{\lambda\mu} (\lambda+\mu+a_1){M}_{\lambda\mu}+\sum_{p=2}^\infty\,\frac{a_p \N^{-p+1}}{(p!)^2} \,\Tr \, ({M}^\dagger(t)M(t))^p\,,\label{HamiltonianCdef}
\end{equation}
and the dot product ‘‘$\cdot$'' is defined as:
$
A \cdot B:= \int \extd t\, \sum_{i,j=1}^N\, A_{ij}(t) B_{ij}(t)\,.
$
The resulting Boltzmann type partition function must be rewritten in a clever way introducing the two momenta, positives in the large $\N$:
\begin{equation}
p_1:= \lambda+2\sigma\,, \qquad p_2:= \mu+2\sigma\,,
\end{equation}
together with the mass: $m:=a_1-4\sigma$, such that the propagator becomes:
$
C_{p_1p_2}:=E^{-1}(p_1,p_2)\,,
$
where $E(p_1,p_2):=p_1+p_2+m$ is the \textit{energy} of the mode $(p_1,p_2)$. Then, in the large $N$ limit, the equilibrium partition function looks like (but differs from) an ordinary Euclidean field theory, with nearly continuous momenta $p$. 
\medskip

Perturbation theory can be investigated using the ordinary Feynman graphs machinery. We materialize vertices of the theory as bipartite (bi)-colored completes graphs, called bubbles, such that black and white noises correspond respectively to fields $M$ and $\overline{M}$, hooked with colored edges materializing Kronecker delta between their indices. As an example:
\begin{equation}
\vcenter{\hbox{\includegraphics[scale=0.7]{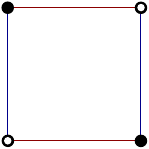}}}\,\equiv \,\sum_{{\color{blue}i},{\color{red}j},{\color{blue}k},{\color{red}l}} M_{{\color{blue}i}{\color{red}j}}\bar{M}_{{\color{blue}k}{\color{red}j}}M_{{\color{blue}k}{\color{red}l}}\bar{M}_{{\color{blue}i}{\color{red}l}}\,.
\end{equation}
A Feynman graph then looks like a (tri) colored bipartite regular graph, with dotted edges hooked between black and white nodes, corresponding to the Wick contractions (see Figure \ref{figFeynmandiag}). It has to be noticed that a Feynman graph for such a theory is more than a set of vertices and edges, but shares an additional structure, the \textit{faces}, for which we recall the definition:

\begin{definition}
A face is an alternate cycle made of dashed and solid edges. It can be closed or open, and closed faces, corresponding to a complete trace, share a factor $\N$.
\end{definition}\label{defface}
Then, a typical Feynman amplitude $A(G)$, depending on some Feynman graph $G$ scale as $A(G)\sim N^{\omega(G)}$, with:
\begin{equation}
\omega(G)=-\sum_k (k-1) V_k(G)+F(G)
\end{equation}
where $ V_k(G)$ denotes the number of vertices of type ‘‘$2k$'', and $F(G)$ the number of (closed) faces. 
\medskip

\begin{figure}
\begin{center}
\includegraphics[scale=1]{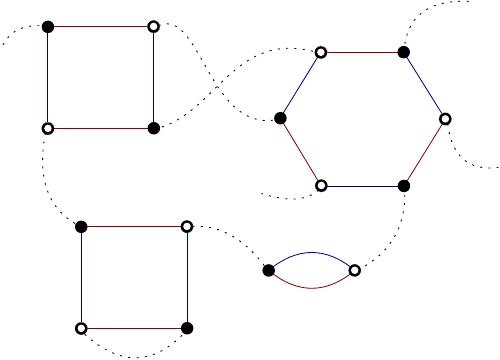}
\end{center}
\caption{A typical Feynman graph involving four external edges, two quartic vertices, a sextic, and a quadratic vertex.}
\end{figure}

We considered two different strategies for constructing renormalization group flow, which we called respectively ‘‘passive scheme'' or ‘‘scaling 1'' and ‘‘active scheme'' or ‘‘scaling 2''. For scaling 1, we consider the $\Lambda$ dependent family of models:
\begin{equation}
Z_{\text{eq}}^{(\mathbb{C},\Lambda)}[L]:=\int \extd M \extd\overline{M} e^{-\sum_{p_1,p_2} \overline{M}_{p_1p_2} (C^{(\Lambda)})^{-1}_{p_1p_2} M_{p_1p_2}-U_{\text{int}}[M,\overline{M}]+\overline{L}\cdot M+\overline{M}\cdot L}\,,\label{functionpartLambda}
\end{equation}
where $U_{\text{int}}[M,\overline{M}]:=\sum_{n=2}^\infty [a_n \N^{-n+1}/(n!)^2]\, \Tr (M^\dagger M)^n$, and the propagator $C^{(\Lambda)}$ reads:
\begin{equation}
C_{p_1p_2}^{(\Lambda)}:=\frac{1}{2}\frac{\chi_\Lambda(p_1)+\chi_\Lambda(p_2)}{p_1+p_2+m}\,,\label{regularization1}
\end{equation}
where $\chi_\Lambda(x)$ is the windows function:
\begin{equation}
\chi_\Lambda(x):=\theta(4\sigma-x)-\theta(\Lambda-x)\,,
\end{equation}
which is equal to 1 inside the interval $(\Lambda,4\sigma)$ and 0 outside. As explained in \cite{lahoche20241} it is suitable to view $\theta(x)$ as the limit of a sequence of smooth functions $\theta_n(x)$, which converge weakly on the Schwartz space toward the step function $\theta(x)$ as $n\to \infty$, and we assume to take implicitly the limit after the limit $\N\to \infty$ to replace sums by integrals. For instance:
\begin{equation}
\theta_n(x):=\sqrt{\frac{n}{\pi}}\int_0^x \extd y\,e^{-n y^2}\,.\label{regultheta}
\end{equation}
Then the effective average action:
\begin{equation}
\Gamma_\Lambda[\Phi]+\overline{\Phi}\cdot R_\Lambda \cdot \Phi:=\overline{L}\cdot \Phi+\overline{\Phi}\cdot L-\ln Z_{\text{eq}}[L]\,,\label{GammaTrue}
\end{equation}
where $R_\Lambda:=(C^{(\Lambda)})^{-1}$, interpolates between $U_{\text{int}}$ and:
\begin{equation}
\Gamma_{\Lambda=0}[\Phi] \equiv \Gamma[\Phi]-\sum_{p_1,p_2} \, \bar{\Phi}_{p_1p_2} (p_1+p_2+m){\Phi}_{p_1p_2}\,,
\end{equation}
where $\Gamma[\Phi]$ is the full effective action (including all quantum corrections) and in these equations ${\Phi}_{p_1p_2}$ is the \textit{classical field}:
\begin{equation}
\Phi:= \frac{\delta }{\delta \overline{L}}\ln Z_{\text{eq}}[L]\,.
\end{equation}
The flow equation can be deduced by taking the derivative with respect to $\Lambda$, and we get:
\begin{equation}
\boxed{\partial_\Lambda \Gamma_\Lambda[\Phi]=\left( \frac{\delta}{\delta L}\cdot \partial_\Lambda  (C^{(\Lambda)})^{-1} \cdot \frac{\delta}{\delta \overline{L}} \right) W_\Lambda\,,}\label{Wett}
\end{equation}
where $W_\Lambda:=\ln Z_{\text{eq}}[L]$. The local truncation, at the leading order of the derivative expansion, is then defined by the expansion:
\begin{equation}
\Gamma_\Lambda[\Phi]=2z(\Lambda)\sum_{p_1,p_2}\bar{\Phi}_{p_1p_2} (p_1+p_2+\bar{\mu}_2(\Lambda))\Phi_{p_2p_1}+\sum_{k>1} \frac{\mu_{2k}(\Lambda)}{k!^2 \N^{k-1}} \Tr ({\Phi}^\dagger \Phi)^k\,,\label{truncation}
\end{equation}
where from the initial condition $z(4\sigma)=0$. The flow equations can be computed by taking the functional derivative with respect to the classical field. For $\mu_2$ for instance, we get: 
\begin{equation}
\partial_\Lambda \mu_2=-\frac{\mu_4}{4\N} \sum_p (p+m)  \frac{\delta_n(\Lambda-p)}{((z+1)p+(\mu_2+m))^2}\,,\label{flowequ2S1}
\end{equation}
or, introducing the renormalized dimensionless couplings:
\begin{equation}
\tilde{m}:=\Lambda^{-1} m\,,\quad \tilde{\mu}_2:=(1+z)^{-1}\Lambda^{-1} ({\mu}_2+m)\,,\quad \tilde{\mu}_4:=(1+z)^{-3}\Lambda^{-1} \tilde{\rho}(\Lambda) {\mu}_4\,,\label{defdimensionless}
\end{equation}
we get:
\begin{equation}
\boxed{\Lambda\partial_\Lambda \tilde{\mu}_2=-(1+\eta)\tilde{\mu}_2-\frac{\tilde{\mu}_4}{4} \frac{1+\tilde{m}}{(1+\tilde{\mu}_2)^2}\,,}\label{flowequationmu2}
\end{equation}
In ‘‘scheme 2'', we add in the  classical Hamiltonian, the  regulator such that, 
\begin{equation}
H_\infty^{\mathbb{C}} \to H_\infty^{\mathbb{C}}+\Delta S_k,\label{regulatedH}
\end{equation}
in order to interpolate with the classical Hamiltonian (for $k\to \infty$) and the effective action $\Gamma$ as $k\to 0$. we impose to the regulator the following structure:
\begin{equation}
R_k(p_1,p_2)=Z(k)k\left(\frac{4\sigma}{4\sigma-k}\right)\left(2-\frac{p_1+p_2}{k}\right) f\left(\frac{p_1}{k}\right)f\left(\frac{p_2}{k}\right)\label{eqR}
\end{equation}
where the function $f(x)$ is inspired from the Litim's one:
\begin{equation}
f(x):=\theta(1-x)\,,
\end{equation}
and $Z$ is the wave function renormalization. Note that this regulator is something like a mixing between the standard Litim regulator and the Callan-Symanzik regulator \cite{otto2022regulator}. The equation describing how the effective action $\Gamma_k$ changes as $k$ changes is the so-called \textit{Wetterich-Morris} equation \cite{Berges_2002,Delamotte_2012}, which reads for complex matrices:
\begin{equation}
\boxed{\dot{\Gamma}_k=\sum_{P=(p_1,p_2)}\,  \frac{\dot{R}_k(P)}{\Gamma^{(2)}_k(P)+R_k(P)}\,,}\label{Wett2}
\end{equation}
where 
\begin{equation}
\dot{X}:=k\frac{\extd X}{\extd k}\,.
\end{equation}
In this scheme, and at the leading order of the derivative expansion, it is suitable to make use of the following ansatz:
\begin{equation}
\Gamma_k [\overline{M},M]= Z(k) \sum_{p_1,p_2} \overline{\Phi}_{p_1p_2}(p_1+p_2+2u_2(k)) \Phi_{p_1p_2}+ \sum_{n>1} \frac{u_{2n}(k)}{(n!)^2 \N^{n-1}} \Tr (\Phi^\dagger \Phi)^n\,.\label{truncationPhi2}
\end{equation}
From which, we easily deduce the flow equation for the dimensionless couplings:
\begin{equation}
\bar{u}_{2n}=Z^{-n}(k) k^{-\frac{3-n}{2}} u_{2n} \,.
\end{equation}
Explicitly, we get:
\begin{equation}
\boxed{\dot{\tilde{u}}_2=-(1+\tilde{\nu})\tilde{u}_2-\frac{\tilde{u}_{4}}{8\pi} \frac{J_{\tilde{\nu}}(k)}{(1+\tilde{u}_2)^2}\,,}\label{equationflowu2S2}
\end{equation}

\begin{equation}
\boxed{\dot{\tilde{u}}_4=-2\left(1+\nu\right)\tilde{u}_4- \frac{\tilde{u}_6}{6\pi} \frac{J_{\tilde{\nu}}}{(1+\tilde{u}_2)^2}+ \frac{\tilde{u}_4^2}{4\pi} \frac{J_{\tilde{\nu}}}{(1+\tilde{u}_2)^3}\,,}\label{flowequationu4tilde}
\end{equation}
with:
\begin{equation}
J_{\tilde{\nu}}(k)=\frac{1}{2\tilde{\sigma}^2} \left[\sqrt{4\tilde{\sigma}-1}+\int_0^1 \extd q \sqrt{q(4\tilde{\sigma}-q)}\left(\tilde{\nu}(2-q)+2\right)\right]\,.\label{defJ}
\end{equation}
\begin{remark}
Note that, except for the anomalous dimension, which depends spuriously on the truncation, equation \eqref{equationflowu2S2} is indeed true beyond local expansion, as the IR cut-off $k$ is small enough \cite{blaizot2006nonperturbative}. 
\end{remark}
\section{Evidences for a discontinuous phase transition}\label{seceqscheme2}

\begin{figure}[H]
\begin{center}
\includegraphics[scale=0.7]{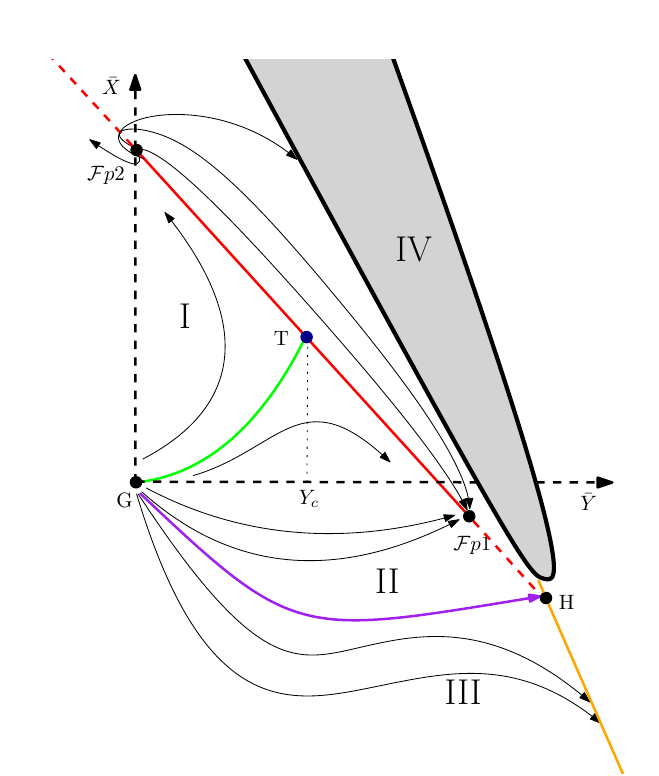}
\end{center}
\caption{Qualitative illustration of the different phases. In the region in the interior of the thick black line, the denominator of $\tilde{\nu}$ is negative.}\label{figFlow}
\end{figure}

In the rest of this paper, we essentially focus on scheme 2, and we will briefly summarize the evidence in favor of a first-order phase transition. For a sextic truncation, we get essentially two reliable fixed points in the deep IR, which we call $fp1$ and $fp2$, and which are respectively purely attractive and purely repulsive. Figure \ref{figFlow} shows qualitatively the behavior of the RG flow for the sextic truncation, in the plane spanned by the two fixed points and the Gaussian one. Note that in the deep IR,  the scale dependence of the canonical dimension can be neglected. The flow reveals different phases, that we call $I$, $II$, and $III$ on the same figure. Coordinates $X,Y$ are defined as:
\begin{align}
\bar{X}&:=-0.03\tilde{u}_2+0.17 \tilde{u}_4-0.98 \tilde{u}_6\\
\bar{Y}&:=-0.59 \tilde{u}_2 + 0.80 \tilde{u}_4 +0.16 \tilde{u}_6.
\end{align}

\begin{figure}[H]
\begin{center}
\includegraphics[scale=0.8]{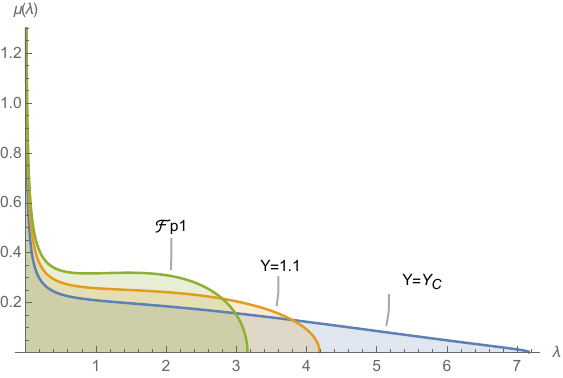}\quad \includegraphics[scale=0.73]{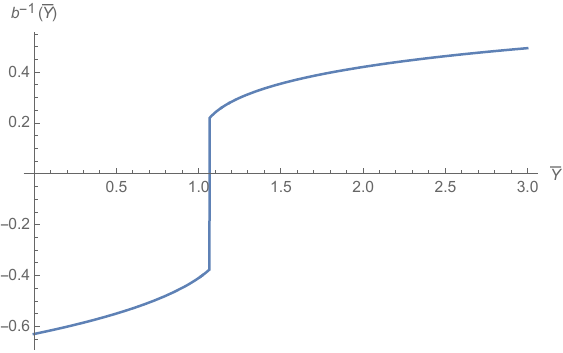}
\end{center}
\caption{On the left: Evolution of the effective eigenvalue distribution along the red curve. On the right: Behavior of the inverse of the solution for the edge boundary with $\bar{Y}$, along the red curve. $\bar{Y}_c \approx 1.0135$.}\label{figflowIR01Prime}
\end{figure}

The first order phase transition occurs along the solid red lines joining the two fixed points, at the point $T$ with coordinate $Y_c\approx 1.1035$. We choose for order parameter the properties of the eigenvalue distribution of the matrix $\chi=M^\dagger M$, assumed to be confined in the interval $[a,b]$ (see Appendix \ref{App6}). Relevant solutions require $a=0$, and the results, along the red curve, are summarized in Figure \ref{figflowIR01Prime}. The Figure shows that, as we move toward $fp2$, the eigenvalue distribution spreads out, until a critical value $Y_c$, at which no finite solution for $b$ exists.  Hence, along this curve, $b$ spontaneously jumps toward a finite value and plays the role of the (discontinuous) order parameter.

\section{Beyond vertex expansion: closing hierarchy at equilibrium}\label{closing1}

In the previous section, we considered both derivative expansion and vertex expansion, by disregarding the role of symmetries and Ward identities. Taking into account the relations coming from Ward identities in the planar sector (because we implicitly consider the large $\N$ limit), we will:
\begin{enumerate}
    \item Improve the computation of the anomalous dimension, taking into account the intrinsic external momenta dependency of the quartic vertex.
    \item Close the hierarchy, expressing the effective local sextic in terms of the quartic and quadratic ones. 
\end{enumerate}
Note that this technique has been also considered in the context of tensorial field theory, in the leading sector (melonic), and beyond \cite{Lahoche:2018oeo}. 

\subsection{Symmetries and modified planar Ward identities}\label{planarWI}

let's look back at the definition of
 the large $\N$ equilibrium partition function \ref{parteqcomplex2}, with the Hamiltonian \ref{HamiltonianCdef}, that reads in terms of the $p$ indices:
\begin{equation}
H_\infty^{\mathbb{C}}[M,\overline{M}]:= \sum_{p_1,p_2} \overline{M}_{p_1p_2} E_k(p_1,p_2){M}_{p_1p_2}+\sum_{q=2}^\infty\,\frac{a_q \N^{-q+1}}{(q!)^2} \,\Tr \, ({M}^\dagger(t)M(t))^q\,,\label{HamiltonianCdef2}
\end{equation}
where the Hamiltonian includes regularization, like for instance in \ref{regulatedH}, blinded in the kinetic kernel $E_k(p_1,p_2)$ (the parameter  $k$ plays the role of IR cut-off as the UV  cut-off and this parameter occurs in equation \ref{regulatedH} previously).
\medskip 

Obviously, the Hamiltonian \eqref{HamiltonianCdef2} is not invariant under $\mathcal{U}(\N)^2$ transformations: $M\to U M V$, with $U,V$ some $\N \times \N$ unitary matrices. However, the partition function \ref{parteqcomplex2} is invariant by construction, because we integrate over all matrices, and the Jacobian of the transformation is: 
\begin{eqnarray}
J = \vert \det (U\otimes V) \vert^2=1\,.
\end{eqnarray}
We then have:
\begin{equation}
0=\int \extd M \extd \overline{M}\, \left(e^{- H_\infty^{\mathbb{C}}[M,\overline{M}]+ \overline{L}\cdot M+\overline{M}\cdot L}-e^{- H_\infty^{\mathbb{C}}[UMV,\overline{U}\,\overline{M}\,\overline{V}]+ \overline{L}\cdot UMV+\overline{U}\,\overline{M}\,\overline{V}\cdot L}\right)\,.\label{identityvar1}
\end{equation}
Because the $\mathcal{U}(\N)^2$ transformations acting on the left and on the right are independent, we can consider infinitesimal transformations  "on the left" or "on the right". For instance, we can consider the left transformation given by:
\begin{equation}
V=I\,,\qquad U=I+i\epsilon\,,
\end{equation}
where $\epsilon$ is an infinitesimal Hermitian matrix: $\epsilon=\epsilon^\dagger$. The interaction term is invariant by construction, and because the integration measure is also invariant, the variation of the partition function then comes to the kinetic and the source terms. Let us compute them separately.

\paragraph{Source term.} The variation of the source is easy to compute from the definition, we have for instance at first order in $\epsilon$
\begin{align}
\overline{U}\,\overline{M}\,\overline{V}\cdot L-\overline{M}\cdot L&=-i \left(\overline{\epsilon}\,\overline{M}\right)\cdot L\\
&= -i \sum_{p_1,p_2,p_3} \overline{\epsilon}_{p_1p_2} \overline{M}_{p_2p_3} L_{p_1p_3}\\
&=-i \sum_{p_1,p_2,p_3}  \overline{M}_{p_2p_3} L_{p_1p_3} {\epsilon}_{p_2p_1}\\
&=-i \sum_{p_1,p_2,p_3}  \overline{M}_{p_1p_3} L_{p_2p_3} {\epsilon}_{p_1p_2}\,.
\end{align}
Such that the variation of the source term is finally:
\begin{equation}
\delta_\epsilon \left(\overline{L}\cdot M+\overline{M}\cdot L\right)= i \sum_{p_1,p_2,p_3} \left({M}_{p_2p_3} \overline{L}_{p_1p_3}- \overline{M}_{p_1p_3} L_{p_2p_3}\right){\epsilon}_{p_1p_2}\,.
\end{equation}
Where we introduced the symbol $‘‘\delta_\epsilon"$ as the variation at order $1$ in $\epsilon$. 

\begin{figure}[H]
\begin{center}
\includegraphics[scale=1.3]{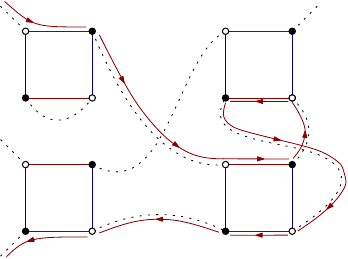}
\end{center}
\caption{A typical Feynman graph with $4$ vertices, $4$ external edges, and $6$ internal edges. The red path with arrows materializes a typical external face.}\label{figFeynmandiag}
\end{figure}

\paragraph{Kinetic term.} In the same way and in first order, the variation of the kinetic term reads:
\begin{align}
\delta_\epsilon \left(\sum_{p_1,p_2} \overline{M}_{p_1p_2} E_k(p_1,p_2){M}_{p_1p_2}\right)&=i \sum_{p_1,p_2,p_1^\prime} \Big( - \overline{\epsilon}_{p^\prime_1p_1} \overline{M}_{p_1p_2} E_k(p_1^\prime,p_2){M}_{p_1^\prime p_2}\\
&\quad +\overline{M}_{p_1p_2} E_k(p_1,p_2) \epsilon_{p_1p^\prime_1}{M}_{p_1^\prime p_2}\Big)\,,
\end{align}
or, because $\overline{\epsilon}_{p_1^\prime p_1}=\epsilon_{p_1 p_1^\prime}$,
\begin{equation}
\delta_\epsilon \left(\sum_{p_1,p_2} \overline{M}_{p_1p_2} E_k(p_1,p_2){M}_{p_1p_2}\right)=i \sum_{p_1,p_2,p_1^\prime} \overline{M}_{p_1p_2}\left(E_k(p_1,p_2)-E_k(p_1^\prime,p_2)  \right) {M}_{p_1^\prime p_2} \epsilon_{p_1p^\prime_1}\,.
\end{equation}
Then, the computation of the variation of the identity \eqref{identityvar1} at first order leads, because the cancellation has to be independent of $\epsilon$:
\begin{proposition}
The partition function obeys the following differential equation (\textbf{Ward identity}):
\begin{align}
\sum_{p_2} \Bigg( \delta E_k(p_1,p_1^\prime,p_2) \frac{\partial^2}{\partial L_{p_1p_2}\partial \overline{L}_{p_1^\prime p_2}} + \overline{L}_{p_1p_2} \frac{\partial}{\partial \overline{L}_{p_1^\prime p_2}}-{L}_{p_1^\prime p_2} \frac{\partial}{\partial {L}_{p_1 p_2}}   \Bigg)Z_{\text{eq}}^\mathbb{C}[L,\overline{L}]=0\,,\label{WardId}
\end{align}
where:
\begin{align}
\delta E_k(p_1,p_1^\prime,p_2) :&=E_k(p_1^\prime,p_2)-E_k(p_1,p_2)\\
&\equiv (p_1^\prime-p_1)+R_k(p_1^\prime,p_2)-R_k(p_1,p_2)\,.
\end{align}
\end{proposition}
The previous Ward identity can be easily expressed in terms of the connected correlation functions. Indeed, $Z_{\text{eq}}^\mathbb{C}=: \exp \left(+W_{\text{eq}}^\mathbb{C}\right)$, and the connected correlations function are the $n$-th derivatives of the free energy $W_{\text{eq}}^\mathbb{C}$ with respect to the source field. To write down explicit expressions, we introduce the notation $\vec{p}=:(p_1,p_2)$ for the pair of indices shared by the source field $L_{p_1p_2}\equiv L_{\vec{p}}$. We furthermore make use of the norm:
\begin{equation}
\vert \vec{p}\,\vert :=p_1+p_2\,.
\end{equation}
The connected correlation functions are then defined as:
\begin{equation}
(G_c^{(2n)})_{\vec{p}_1\cdots \vec{p}_{n}, \vec{\bar{p}}_{n+1}\cdots \vec{\bar{p}}_{2n}}= \frac{\partial^{n}}{\partial L_{\vec{p}_1}\cdots L_{\vec{p}_{n}}}\frac{\partial^{n}}{\partial \bar{L}_{\vec{\bar{p}}_{n+1}}\cdots \bar{L}_{\vec{\bar{p}}_{2n}}}\,W_{\text{eq}}^\mathbb{C}
\end{equation}
Note that odd correlation functions vanish by construction in the \textit{symmetric phase}, where the classical field
\begin{equation}
\Phi_{\vec{p}}= \frac{\partial W_{\text{eq}}^\mathbb{C}}{\partial \bar{L}_{\vec{p}}}
\end{equation}
vanishes. More precisely, we define the symmetric phase as follows \cite{Lahoche:2018oeo}:
\begin{definition}
We call the symmetric phase the region of the phase space where an expansion of the classical action $\Gamma_k$ in the power of field is convergent enough. In this region, all the odd vertex functions vanish (including the classical field itself), and the two-point function $G_{k,\vec{p}\,\vec{q}}^{(2)}$ is diagonal:
\begin{equation}
G_{k,\vec{p}\,\vec{q}}^{(2)}=:G_k(\vec{p}\,) \delta_{\vec{p}\,\vec{q}}\,.
\end{equation}
\end{definition}
The Ward identity \eqref{WardId} provides an infinite number of relations between observables, that we can obtain by taking successive derivatives with respect to the classical field $\Phi$. Note however that we have to impose the symmetric phase condition at the end, after the computation of all the derivatives. Finally, it is suitable to rewrite the Ward identities in terms of the 1PI vertex functions instead of connected generating functions, and we rewrite the Ward identity \eqref{WardId} as:
\begin{align}
\sum_{p_2} \Bigg(\delta E_k(p_1,p_1^\prime,p_2) \left(G^{(2)}_{k,\vec{p}\vec{p}\,^\prime}+\overline{\Phi}_{\vec{p}} \Phi_{\vec{p}\,^\prime}\right) + \overline{L}_{p_1p_2} {\Phi}_{\vec{p}\,^\prime}-{L}_{p_1^\prime p_2} \overline{\Phi}_{\vec{p}}   \Bigg)=0\,,\label{WardId2}
\end{align}
where $\vec{p}:=(p_1,p_2)$ and $\vec{p}\,^\prime:=(p_1^\prime,p_2)$, and:
\begin{equation}
G^{(2)}_{k,\vec{p}\vec{p}\,^\prime}:= \left(\Gamma_k^{(2)}+R_k \mathrm{I}\right)^{-1}_{\vec{p}\vec{p}\,^\prime}\,,
\end{equation}
where $\mathrm{I}$ is the identity matrix in the momenta space. In the rest of this section, we need only two Ward identities, obtained by taking respectively the second and fourth functional derivative of \eqref{WardId2} with respect to the classical fields $\Phi$ and $\overline{\Phi}$. Taking the second derivative, and imposing at the end the symmetric phase conditions, we get after a straightforward computation (see for instance \cite{Lahoche:2018oeo,samary2014closed}):
\begin{align}
-&\sum_{p_2} \delta E_k(p_1,p_1^\prime,p_2) G^{(2)}_{k,\vec{p}\,\vec{r}}\,\Gamma_{k,\vec{q}\,\vec{q}\,^\prime\,\vec{r}\,\vec{s}}^{(4)}\,G^{(2)}_{k,\vec{s}\,\vec{p}\,^\prime} + \sum_{p_2}  \Big[\delta E_k(p_1,p_1^\prime,p_2)\delta_{\vec{p}\,\vec{q}}\,\delta_{\vec{p}\,^\prime\,\vec{q}\,^\prime}\\
&+\left(\Gamma_k^{(2)}(\vec{p}\,)+R_k(\vec{p}\,)\right)\delta_{\vec{p}\,\vec{q}}\,\delta_{\vec{p}\,^\prime\vec{q}\,^\prime}-\left(\Gamma_k^{(2)}(\vec{p}\,^\prime\,)+R_k(\vec{p}\,^\prime\,)\right)\delta_{\vec{p}\, \vec{q}}\,\delta_{\vec{p}\,^\prime\vec{q}\,^\prime}\Big]=0\,,\label{WardId3}
\end{align}
where $\vec{q},\vec{q}\,^\prime$ are the external momenta i.e. the momenta of the variables with respect to which we take the derivative. Furthermore, contiguous repeated vector indices assumed are summed, and we made use of the identity:
\begin{equation}
\frac{\partial \overline{L}_{\vec{p}}}{\partial \overline{\Phi}_{\vec{q}}}= \frac{\partial^2 \Gamma_k}{\partial \overline{\Phi}_{\vec{q}}\,\partial{\Phi}_{\vec{p}}}+R_k(\vec{p}\,)\delta_{\vec{p}\,\vec{q}}=:\left(\Gamma_k^{(2)}(\vec{p}\,)+R_k(\vec{p}\,)\right)\delta_{\vec{p}\,\vec{q}}\,.
\end{equation}
We focus on the planar approximation, in the symmetric phase, which imposes some conditions on the effective vertices $\Gamma_k^{(2n)}$. In the perturbative point of view, the effective vertices admit an expansion indexed by 1PI Feynman diagrams i.e. a set of vertices, edges, and faces (see definition \ref{defface}), such that edges correspond to Wick contractions (see Figure \ref{figFeynmandiag}). The edges can be contracted according to the  following rule:
\begin{enumerate}
    \item We cancel all the edges (including the dotted one) between the two boundary nodes (black and white), and we cancel them also.
    \item We merge the remaining lines (initially attached to the white and black nodes), according to their respective colors.
\end{enumerate}

\begin{figure}[H]
\begin{center}
\includegraphics[scale=0.8]{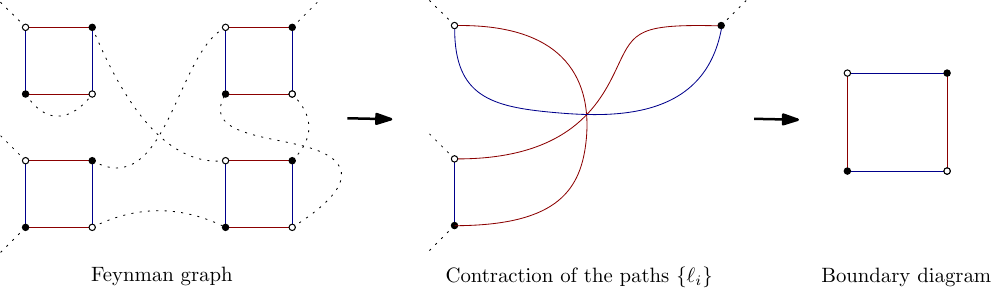}
\end{center}
\caption{Construction of the boundary diagram of some Feynman graph.}\label{figBoundary}
\end{figure}

We furthermore define the concept of boundary diagram. Let $\mathcal{G}$ be a 1PI Feynman diagram involving $V>1$ vertices and $E>1$ external half edges. Let $E_0$ the set of internal edges along the boundary of external faces (see definition \ref{defface}) and let $P=\{\ell_i\}$ be the set of open cycles $\ell_i$ involving dotted edges and edges of color $\ell_i$, bounded by external nodes (hooked to external edges) and such that:
\begin{equation}
\bigcup_{i}\ell_i=E_0\,.
\end{equation}
Then, we have the definition:
\begin{definition}
The boundary diagram $\partial \mathcal{G}$ is then the set of external nodes, hooked together by colored edges corresponding to the paths $\ell_i$ where all the dotted edges have to be contracted. The boundary diagram is then a vertex, which can be connected or not. 
\end{definition}
In Figure \ref{figBoundary}, we show an example of the Feynman diagram pictured in Figure \ref{figFeynmandiag}. In practice, the boundary diagram can be obtained by contraction of the first internal line. 

In this section, we focus on the planar sector, which is the relevant one in the large $\N$ limit \cite{Francesco_1995}. The relevant Feynman diagram in this limit is the planar one, whose dual graphs are zero genus. Interestingly for our purpose, in the planar sector, all the boundary diagrams are connected, and we have the following statement:
\begin{lemma}\label{lemma1}
The Feynman expansion for $\Gamma_k^{(2n)}$ can be indexed by planar diagrams having the same boundary diagram, corresponding to the single connected trace invariant vertices.
\end{lemma}
The proof can be done recursively and is given in Appendix \ref{App7}. We introduce the graphical notation:
\begin{equation}
\Gamma_{k,\vec{p}_1\cdots \vec{p}_{2n}}^{(2n)}\equiv \vcenter{\hbox{\includegraphics[scale=1]{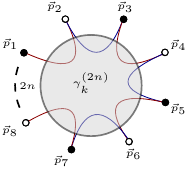}}}+\text{perm}\,,\label{decompgammak}
\end{equation}
where “perm” indicates that we sum all the permutations of external momenta. This notation shows explicitly the structure of the internal paths $\{\ell_i\}$ hooked to external nodes, and the function $\gamma_{k,p_{11},p_{22},\cdots,p_{2n2}}^{(2n)}$ depends only on independent components $(p_{11},p_{22},p_{31},\cdots,p_{2n2})$ (i.e. one momentum per colored edge), the gray disc materializing the formal sum of diagrams. Making use of these notations, the first term of the Ward identity \eqref{WardId3} reads:
\begin{equation}
\sum_{p_2} \delta E_k(p_1,p_1^\prime,p_2) G^{(2)}_{k,\vec{p}\,\vec{r}}\,\Gamma_{k,\vec{q}\,\vec{q}\,^\prime\,\vec{r}\,\vec{s}}^{(4)}\,G^{(2)}_{k,\vec{s}\,\vec{p}\,^\prime} \equiv \vcenter{\hbox{\includegraphics[scale=1]{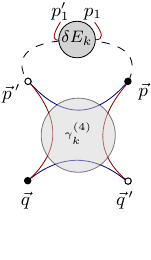}}}\quad + \quad \vcenter{\hbox{\includegraphics[scale=1]{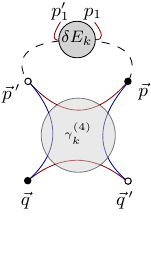}}}
\end{equation}
whereas previously the dotted edges materialized the effective propagator and the little gray circles materialized the operator $\delta E_k$. It is easy to check that the first diagram creates a face (therefore a global factor $\N$) whereas the second one does not. Hence, in the planar approximation, we keep only the first contribution. Finally, let us define the \textit{self energy} $\Sigma_k(\vec{p}\,)$ as:
\begin{equation}
\Gamma_k^{(2)}(\vec{p}\,):=\vert \vec{p}\,\vert - \Sigma_k(\vec{p}\,)\,.
\end{equation}
Then, because of the definition of $\delta E_k$, the previous equation \eqref{WardId3} becomes finally at the leading order:
\begin{equation}
\boxed{\sum_{p_2}\Big[1+\frac{R_k(p_1^\prime,p_2)-R_k(p_1,p_2)}{p_1^\prime-p_1}\Big]G_{k}(\vec{p}\,)\,\gamma_{k,p_2,q_2,p_1^\prime,p_1}^{(4)}\,G_{k}(\vec{p}\,^\prime)-\frac{\Sigma_k(\vec{p}\,^\prime)- \Sigma_k(\vec{p}\,)}{p_1^\prime-p_1}\Bigg\vert_{p_2=q_2} =0\,,}\label{WardId4}
\end{equation}
where the following constraint for external momenta is assumed:
\begin{equation}
p_1=q_1\,,\quad p_1^\prime=q_1^\prime\,,\quad q_2=q_2^\prime\,.\label{momentaconfig4pts}
\end{equation}
Equation \eqref{WardId4} is called the \textit{first modified Ward identity} and is a relation between $4$ point functions with the momenta dependency of the 1PI $2$-point function $\Sigma_k(\vec{p}\,)$. 
\medskip

\begin{figure}
\begin{center}
\includegraphics[scale=1]{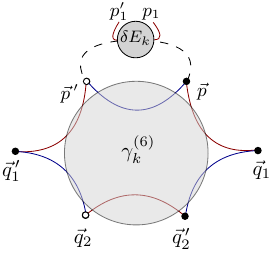}
\end{center}
\caption{Momenta configuration for the computation of equation \eqref{eqWard61}.}\label{Figmomentaconfig}
\end{figure}

We will also need the Ward identity at the next order, relating the vertex functions at $6$ and  $4$-points. To this end, we take the fourth derivative of \eqref{WardId2} two times with respect to $\Phi$ and two times with respect to $\bar{\Phi}$. We denote respectively by $(\vec{q}_1,\vec{q}_2)$ and $(\vec{q}_1^{\,\prime},\vec{q}_2^{\,\prime})$ the external momenta shared by fields $\bar{\Phi}$ and $\Phi$, and we get:
\begin{align}
\nonumber\sum_{p_2} &\Bigg(\delta E_k(p_1,p_1^\prime,p_2) G_{k}(\vec{p}\,^\prime)G_{k}(\vec{p}\,)\left(\gamma_{k,p_2,p_1^\prime,p_1,q^\prime,q,q^{\prime\prime}}^{(6)}-2\gamma^{(4)}_{k,p_2,q^{\prime\prime},p^\prime,q^\prime}\gamma^{(4)}_{k,p_2,q^{\prime\prime},p,q} G_k(p_2,q^{\prime\prime})\right)\\
&-\Gamma^{(4)}_{k,\vec{p} \vec{q_2}^{\prime} \vec{q}_1\vec{q}_2} \delta_{\vec{p}\,^{\,\prime}\vec{q}_1^{\,\prime}}
+ \Gamma^{(4)}_{k,\vec{q_1}^{\prime} \vec{q_2}^{\prime} \vec{p}\,^\prime\,\vec{q}_2} \delta_{\vec{p}\vec{q}_1}
\Bigg)=0\,,\label{eqWard61}
\end{align}
where external momenta are such that:
\begin{equation}
\vec{q}_1^{\,\prime}=(p_1^\prime,q^\prime)\,,\quad \vec{q}_1=(p_1,q)\,,\quad \vec{q}_2=(q^{\prime\prime},q^\prime)\,,\quad \vec{q}_2^{\,\prime}=(q^{\prime\prime},q)\,,\label{SmodifiedWI}
\end{equation}
assuming $p_1\neq p_1^\prime \neq q^{\prime\prime}$; the configuration of momenta for the sextic effective vertex is shown in Figure \ref{Figmomentaconfig}. We call \textit{second modified Ward identity} the equation \eqref{eqWard61}.

\subsection{Local approximation -- Scheme 2}\label{sectionWardS2}
In this section, we investigate the Ward identities in the vicinity of the local potential approximation. This approximation usually focuses on the first order of the derivative expansion and neglects the momenta dependency of the effective vertices. In this case, the non-local structure of the interaction requires considering also the derivative of the effective vertex with respect to the external momenta \cite{Lahoche:2018oeo}. The Ward identities computed in the previous section allow computing the derivative of the vertices. More precisely, we will see that the first and second modified Ward identities introduced in the previous sections provide the full expression for the anomalous dimension in the local sector but also allow closing the hierarchy around the quartic sector. 
\medskip

\subsubsection{First modified Ward identity.} 
\begin{figure}
\begin{center}
\includegraphics[scale=0.4]{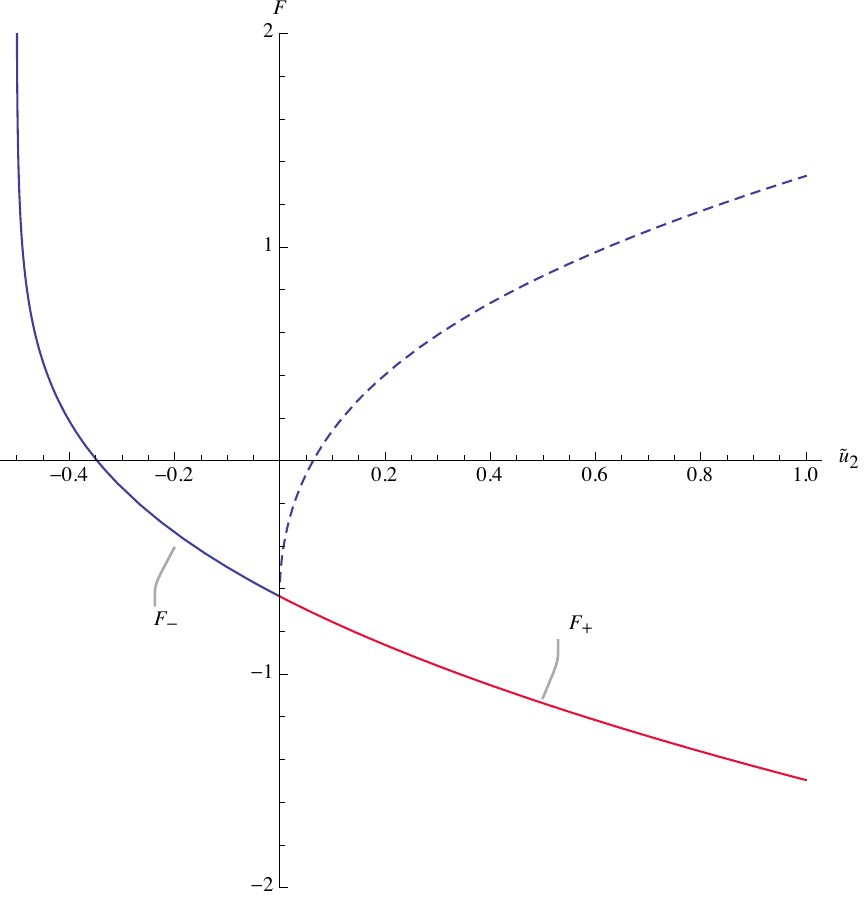}\qquad \qquad \includegraphics[scale=0.4]{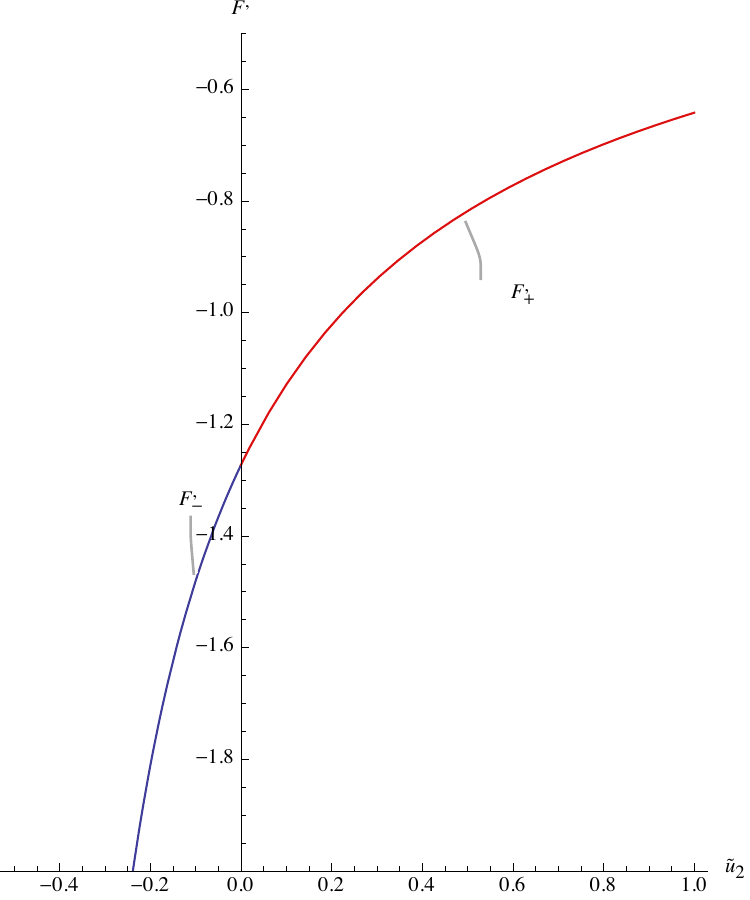}
\end{center}
\caption{On the left: The two branches of solution, $F_{\pm}$. On the left: Continuity of the derivatives $F_{\pm}^\prime$.}\label{solF}
\end{figure}

First, let us consider  the limit $p_1^\prime \to p_1\to 0$ in equation \eqref{WardId4}. Then by  setting $q_2=0$ we have:
\begin{equation}
\N\int \rho(q) \left[1+\frac{\extd R_k}{\extd p_1}(0,q)\right] G_{k}^2(0,q) \gamma_{k,q000}^{(4)}+(\tilde{Z}(k)-1)=0\,,\label{FMWI0M}
\end{equation}
where we use the definition of $\tilde{Z}(k)$, and we replace the discrete sum with a continuous integral involving the asymptotic distribution $\rho(q)$. We then use the modified Litim's regulator, as defined in \eqref{eqR}. This in particular implies that the windows of momenta allowed by the derivative with respect to $p_1$ in the previous equation are the same as the windows of momenta allowed by $\dot{R}_k$ in the flow equation, and the approximations used in the second case has then to be valid in the first case. Hence, because we focus on the local approximation, it makes sense to replace $\gamma_{k,q000}^{(4)}$ by $\gamma_{k,0000}^{(4)}$:
\begin{equation}
\int \rho(q) \frac{\extd R_k}{\extd p_1}(0,q)\, G_{k}^2(0,q) \gamma_{k,q000}^{(4)}\approx  \frac{1}{2}\gamma_{k,0000}^{(4)}\times\int \rho(q) \frac{\extd R_k}{\extd p_1}(0,q)\, G_{k}^2(0,q)\,.
\end{equation}
This approximation can be also motivated, without the local assumption, from the observation that, for $k$ small enough, the allowed windows of momenta selected by the regulator are closed to $q=0$. For the first term on the left-hand side, this approximation is difficult to motivate because far enough from the IR regime, the power counting can be very different, especially using scheme 1 for the scaling. Note however that the flow equation for $\Gamma_k^{(2)}$ reads:
\begin{equation}
\dot{\Gamma}_k^{(2)}(\vec{p}\,)=-\sum_{\vec{q}} \Gamma_{k,\vec{p}\,\vec{p}\,\vec{q}\,\vec{q}}^{(4)}\, G^2(\vec{q}\,) \dot{R}_k(\vec{q}\,)\,.
\end{equation}
Once again, if $k$ is small enough, the $\vec{q}$ dependency of $\Gamma_{k,\vec{p}\,\vec{p}\,\vec{q}\,\vec{q}}^{(4)}$ can be neglected, setting $\vec{q}=\vec{0}$ as the selected windows of momenta is closed to zero. Then:
\begin{equation}
\dot{\Gamma}_k^{(2)}(\vec{p}\,)\approx - \Gamma_{k,\vec{p}\,\vec{p}\,\vec{0}\,\vec{0}}^{(4)}\,\N \mathcal{L}_2(k)\,,
\end{equation}
with the definition:
\begin{equation}
 \N\mathcal{L}_n(k):=\sum_{q} \, G^n(0,q) \dot{R}_k(0,q)\,.
\end{equation}
Hence, we have:
\begin{align}
\nonumber G_{k}^2(0,q) \gamma_{k,q000}^{(4)}&=-\frac{1}{\N\mathcal{L}_2}G_{k}^2(0,q) \dot{\Gamma}_k^{(2)}(0,q)\\
&=-\frac{1}{\N\mathcal{L}_2} G_{k}^2(0,q) \left(\dot{\Gamma}_k^{(2)}(0,q)+\dot{R}_k(0,q)\right)+\frac{1}{\N\mathcal{L}_2} G_{k}^2(0,q) \dot{R}_k(0,q)\,.\label{methodQ4}
\end{align}
To compute the sum over $q$ of the last term, we can use the truncation to express $G_k$, because the selected windows of momenta are the same as in the flow equations (it is not an additional assumption). The first term furthermore can be rewritten as:
\begin{equation}
- \sum_q G_{k}^2(0,q) \left(\dot{\Gamma}_k^{(2)}(0,q)+\dot{R}_k(0,q)\right)=\frac{\extd }{\extd s}\,  \left(\sum_q G_{k}(0,q)\right)\,,
\end{equation}
where $s:=\ln (k)$. The UV contributions, unaffected by the regulator are essentially the same for $G_{k}(0,q)$ and $G_{k+\extd k}(0,q)$. Hence, we make the crucial assumption that, for $k$ small enough, it makes sense to express $G_k$ with the IR truncation. Explicitly, assuming $k$ is small enough (IR regime):

\begin{align}
\frac{1}{\N}&\,\sum_q G_{k}(0,q)\longrightarrow\,\int_0^{4\sigma} \extd q \,  \frac{\rho(q)}{\tilde{Z} q+2\tilde{Z} u_2+R_k(0,q)}\label{mamaito}\\
&=\frac{1}{\tilde{Z}}\left[\int_0^{k} \extd q \,  \frac{\rho(q)}{2k+ 2u_2}+\int_k^{4\sigma} \extd q \,  \frac{\rho(q)}{ q+2 u_2}\right]\\
& \approx \frac{1}{\tilde{Z}}\Bigg[\underbrace{\frac{\sqrt{k}}{3 \pi  \left({\sigma }\right)^{3/2} \left(1+\tilde{u}_2\right)}+\frac{1}{\sigma}+\frac{\sqrt{k}}{ \sigma^{3/2}} \mathrm{F}_\pm(\tilde{u}_2)+\frac{\tilde{u}_2 k}{\sigma^2} +\mathcal{O}(k^2)}_{:=g(k,\tilde{u}_2)}\Bigg]\,,
\end{align}
where the functions $\mathrm{F}_\pm$ are defined as follows:
\begin{equation}
\mathrm{F}_\pm(\tilde{u}_2):=-\frac{\left(\pm\sqrt{2} \sqrt{\tilde{u}_2} \left(\pi \mp 2 \tan ^{-1}\left(\frac{1}{\sqrt{2} \sqrt{\tilde{u}_2}}\right)\right)+2\right)}{\pi}
\end{equation}
The sign $\pm$ refers to the sign of the mass $\tilde{u}_2$, and the two functions are represented in Figure \ref{solF}. For $\tilde{u}_2>0$, we have two branches of solutions, which have the same limit for $\tilde{u}_2\to 0$:
\begin{equation}
\lim_{\tilde{u}_2\to 0}\, F_{\pm}(\tilde{u}_2)=-\frac{2}{\pi}\approx -0.64\,.
\end{equation}

\begin{figure}
\begin{center}
\includegraphics[scale=0.53]{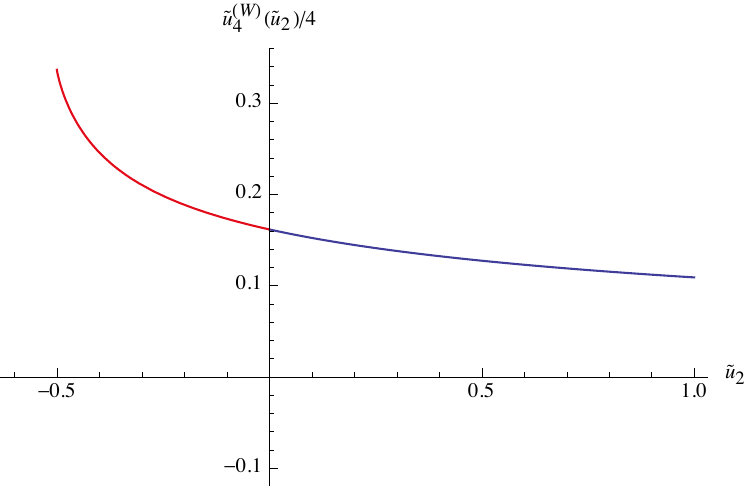}\qquad \quad\includegraphics[scale=0.53]{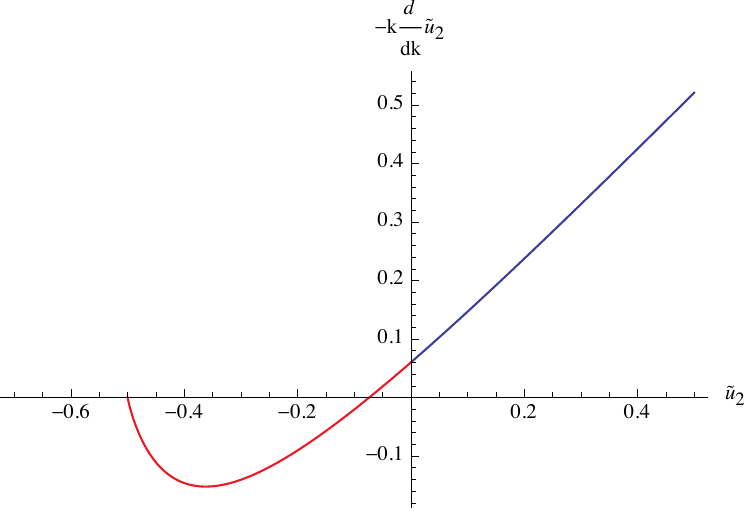}
\end{center}
\caption{On the left: Behavior of the solution for the effective quartic coupling (rescaled by a factor $\tilde{\sigma}^{3/2}$). On the right: Behavior of the beta function for $\tilde{u}_2$.}\label{figu4Ward}
\end{figure}

Now, computing the derivative of the sum \eqref{mamaito} with respect to $s:=\ln(k)$, we get:
\begin{align}
\frac{1}{\N}\frac{\extd }{\extd s}&\,\sum_q G_{k}(0,q)\approx \frac{1}{\tilde{Z}}\Bigg[-\tilde{\nu} g(k,\tilde{u})+k\frac{\partial g}{\partial k}+\frac{\partial g}{\partial \tilde{u}_2}\dot{\tilde{u}}_2\Bigg]\,.
\end{align}
Each term can be easily computed, and using the flow equation \eqref{equationflowu2S2} for $\tilde{u}_2$, we get:
\begin{align}
\nonumber&\frac{1}{\N}\frac{\extd }{\extd s}\,\sum_q G_{k}(0,q)\approx \frac{1}{\tilde{Z}}\Bigg[-\tilde{\nu} g(k,\tilde{u})+ \bigg(\frac{\sqrt{k}}{6 \pi  \left({\sigma }\right)^{3/2} \left(1+\tilde{u}_2\right)}+\frac{\sqrt{k}}{2 \sigma^{3/2}} \mathrm{F}_\pm(\tilde{u}_2)+\frac{\tilde{u}_2 k}{\sigma^2} \bigg) \\
&-\bigg(-\frac{\sqrt{k}}{3 \pi  \left({\sigma }\right)^{3/2} \left(1+\tilde{u}_2\right)^2}+\frac{\sqrt{k}}{\sigma^{3/2}} \mathrm{F}_\pm^\prime(\tilde{u}_2)+\frac{k}{\sigma^2}  \bigg)\bigg((1+\tilde{\nu})\tilde{u}_2+\frac{\tilde{u}_{4}}{8\pi} \frac{J_{\tilde{\nu}}(k)}{(1+\tilde{u}_2)^2}\bigg)\Bigg]\,,\label{equationsum}
\end{align}
where:
\begin{equation}
\mathrm{F}_\pm^\prime(\tilde{u}_2)=-\frac{1}{\pi}\left(\frac{2}{1+2\tilde{u}_2}\pm\frac{\pi \mp 2 \tan ^{-1}\left(\frac{1}{\sqrt{2} \sqrt{\tilde{u}_2}}\right)}{\sqrt{2} \sqrt{\tilde{u}_2}}\right)\,.
\end{equation}
Furthermore, $\mathcal{L}_n(k)$ can be computed exactly:
\begin{equation}
\mathcal{L}_n(k)=\frac{k^{1-n}}{2^n\pi} \tilde{Z}^{1-n} \frac{J_{\tilde{\nu}}}{(1+\tilde{u}_2)^n},\label{definitionLn}
\end{equation}
where $J_{\tilde{\nu}}$ has been defined before in \eqref{defJ}. Moreover, in the deep IR, we have explicitly:
\begin{equation}
J_{\tilde{\nu}}\approx \frac{1}{\tilde{\sigma}^{3/2}} \frac{7}{3}\left(\frac{2 \tilde{\nu}}{5}+1\right)\,,
\end{equation}
and we get finally, at the leading order in $k$:
\begin{align}
\nonumber\frac{1}{4 \N \mathcal{L}_2}\frac{\extd }{\extd s}\,\sum_q G_{k}(0,q)\approx & -\tilde{\nu} \frac{3 \pi (1+\tilde{u}_2)^2}{7\big(1+\frac{2\tilde{\nu}}{5}\big)} \bigg(\bigg(\frac{1}{3\pi (1+\tilde{u}_2)}+F_{\pm}(\tilde{u}_2) \bigg)+\sqrt{\tilde{\sigma}}\bigg)\\\nonumber
&+\frac{3 \pi (1+\tilde{u}_2)^2}{14\big(1+\frac{2\tilde{\nu}}{5}\big)} \bigg(\frac{1}{3\pi (1+\tilde{u}_2)}+F_{\pm}(\tilde{u}_2) \bigg)\\\nonumber
&-\frac{3 \pi (1+\tilde{u}_2)^2}{7\big(1+\frac{2\tilde{\nu}}{5}\big)} \bigg(-\frac{1}{3\pi (1+\tilde{u}_2)^2}+F_{\pm}^\prime(\tilde{u}_2)\bigg)\\
&\quad \times \bigg((1+\tilde{\nu})\tilde{u}_2+\frac{\tilde{u}_{4}}{8\pi} \frac{J_{\tilde{\nu}}(k)}{(1+\tilde{u}_2)^2}\bigg)\,.
\end{align}
Finally,  from the definition of $R_k$:
\begin{equation}
 \frac{\extd R_k}{\extd p_1}(0,q)=-\tilde{Z}(k)\theta(k-q)\,,
\end{equation}
we have, again in the deep IR:
\begin{equation}
\int \rho(q) \frac{\extd R_k}{\extd p_1}(0,q)\, G_{k}^2(0,q)\approx -\frac{2 \left(\frac{1}{\tilde{\sigma }}\right)^{3/2}}{4\times 3 \tilde{Z} k^2\pi}\frac{1}{(1+\tilde{u}_2)^2}\,.
\end{equation}
Furthermore, the truncation implies:
\begin{equation}
 \gamma_{k,0000}^{(4)}=\frac{1}{\N}\tilde{Z}^2(k)k^2 \tilde{u}_4\,,
\end{equation}
and the first modified Ward identity reads finally:
\begin{align}
&\nonumber -\tilde{\nu} \frac{3 \pi (1+\tilde{u}_2)^2}{7\big(1+\frac{2\tilde{\nu}}{5}\big)} \bigg(\bigg(\frac{1}{3\pi (1+\tilde{u}_2)}+F_{\pm}(\tilde{u}_2) \bigg)+\sqrt{\tilde{\sigma}}\bigg)+\frac{3 \pi (1+\tilde{u}_2)^2}{14\big(1+\frac{2\tilde{\nu}}{5}\big)} \bigg(\frac{1}{3\pi (1+\tilde{u}_2)}+F_{\pm}(\tilde{u}_2) \bigg)\\\nonumber
&-\frac{3 \pi (1+\tilde{u}_2)^2}{7\big(1+\frac{2\tilde{\nu}}{5}\big)} \bigg(-\frac{1}{3\pi (1+\tilde{u}_2)^2}+F_{\pm}^\prime(\tilde{u}_2)\bigg)\times \bigg((1+\tilde{\nu})\tilde{u}_2+\frac{\tilde{u}_{4}}{8\pi} \frac{J_{\tilde{\nu}}(k)}{(1+\tilde{u}_2)^2}\bigg)\\
&+\frac{1}{4}-\frac{\tilde{Z} \tilde{u}_4\left(\frac{1}{\tilde{\sigma }}\right)^{3/2}}{12 \pi (1+\tilde{u}_2)^2}+\frac{1}{4}(\tilde{Z}(k)-1)=0
\end{align}
In the deep IR, assuming $\tilde{\nu}=\mathcal{O}(1)$, we have $\tilde{\sigma}\to \infty$, enforcing the condition:
\begin{equation}
\boxed{\tilde{\nu}=0\,\quad \Rightarrow \quad\tilde{Z}(k)=1\,.}
\end{equation}
The fixed point solutions furthermore impose $\tilde{u}_4\sim\tilde{\sigma}^{3/2}$, as we recalled in the previous sections \ref{seceqscheme2}. We then have:
\begin{equation}
\boxed{\tilde{u}_4^{(W)} = \tilde{\sigma}^{3/2}\frac{\frac{3 \pi (1+\tilde{u}_2)^2}{14} \bigg(\frac{1}{3\pi (1+\tilde{u}_2)}+F_{\pm}(\tilde{u}_2) \bigg)-\frac{3 \pi (1+\tilde{u}_2)^2}{7} \bigg(-\frac{1}{3\pi (1+\tilde{u}_2)^2}+F_{\pm}^\prime(\tilde{u}_2)\bigg)\tilde{u}_2+\frac{1}{4}}{\frac{1}{12 \pi (1+\tilde{u}_2)^2}+\frac{1}{8}\bigg(-\frac{1}{3\pi (1+\tilde{u}_2)^2}+F_{\pm}^\prime(\tilde{u}_2)\bigg)}\,.}\label{explicitu4}
\end{equation}
This solution is the effective quartic coupling in the deep IR, and we can show its behavior on the left of Figure \ref{figu4Ward}. The renormalization group flow is finally totally driven by the mass term:
\begin{equation}
\boxed{\beta_2:= \dot{\tilde{u}}_2\vert_{\tilde{u}_4=\tilde{u}_4^{(W)}}=-{\tilde{u}}_2-\frac{7}{24\pi}\,\frac{\tilde{u}_4^{(W)}(\tilde{u}_2)}{(1+{\tilde{u}}_2)^2}\,,}
\end{equation}
and its behavior is shown on the right of Figure \ref{figu4Ward}. A direct inspection shows that it exists one interacting Wilson-Fisher-like fixed point for the value:
\begin{equation}
\tilde{u}_2^{(*)} \approx  -0.07\,,
\end{equation}
Which is relevant, with critical exponents:
\begin{equation}
\theta_-=0.79\,.
\end{equation}
The fixed point is indeed repulsive on both sides, and the flow reaches the positive mass region on the right, and the deep negative mass on the left, until the boundary fixed point at $\tilde{u}_2=-0.5$, where coupling becomes imaginary. 
We call $\text{FPA}$ the fixed point on the right, near the positive mass region. We denote as $\text{FPB}$ the fixed point on the left, on the negative region (which is also the boundary of the symmetric phase region, above this point the solution for the quartic coupling becomes imaginary).

\subsubsection{Next to quartic order.} 
\begin{figure}
 \begin{center}
 \includegraphics[scale=0.5]{u4Ward.pdf}\qquad \includegraphics[scale=0.5]{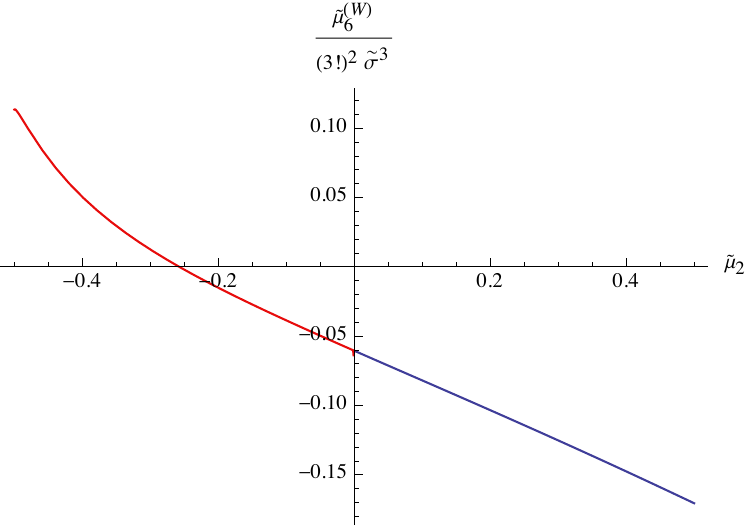}\qquad \includegraphics[scale=0.5]{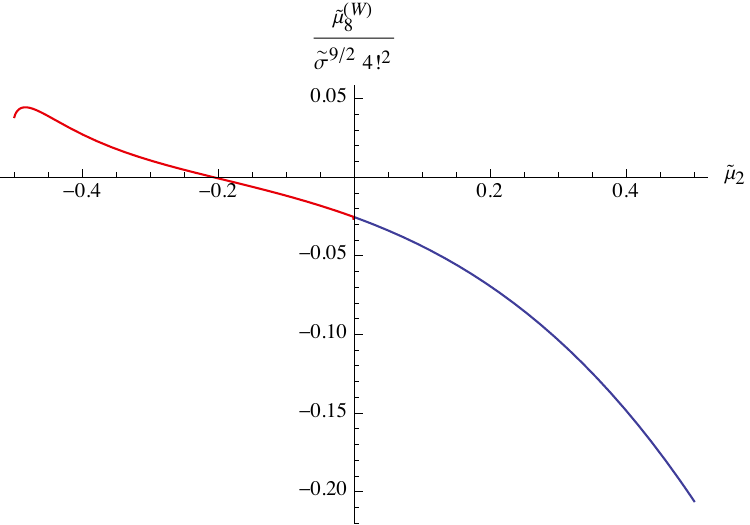}\qquad \includegraphics[scale=0.5]{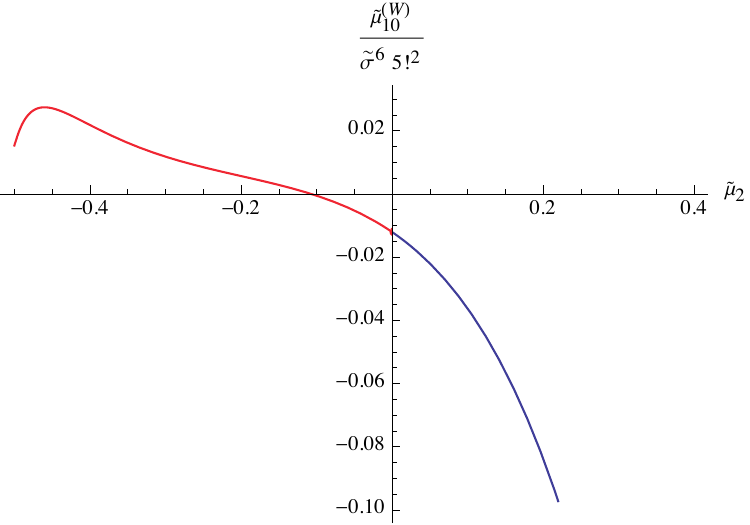}
 \end{center}
 \caption{The constrained solution for $\tilde{u}_{2n}$, rescaled with $\tilde{\sigma}^{3(n-1)/2}$ for $n=1$ to $n=5$.}\label{figu6closed}
\end{figure}

\begin{figure}
\begin{center}
\includegraphics[scale=0.5]{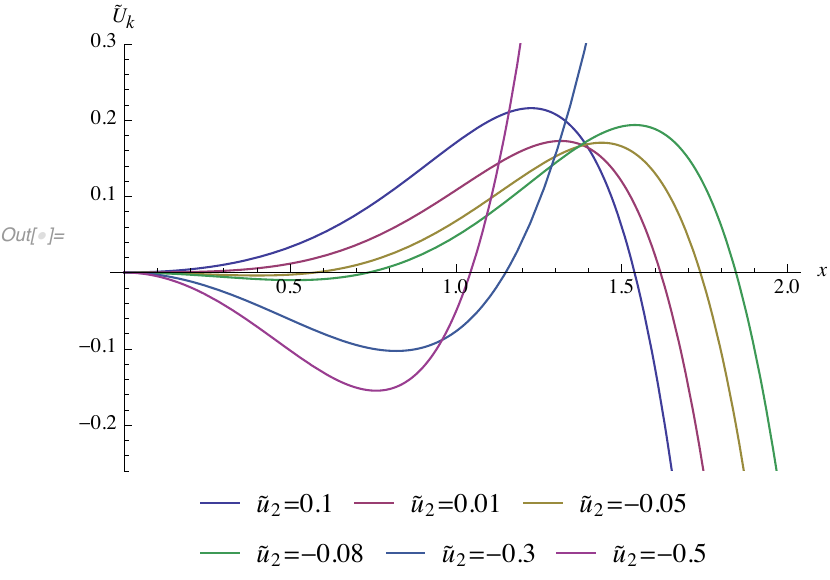}\qquad \includegraphics[scale=0.5]{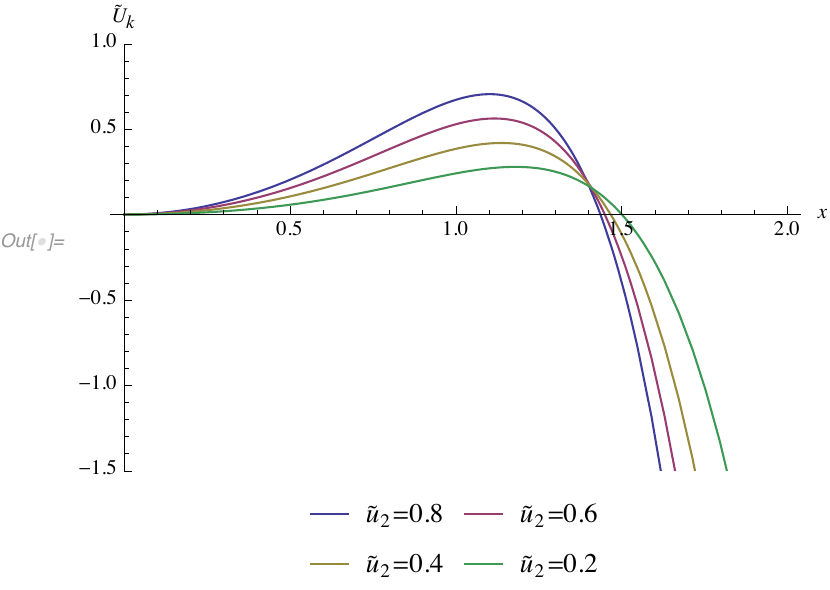}
\end{center}
\caption{Evolution of the effective potential from $\text{FPB}$, along the mass axis. The 
abscissa $x$ is a typical eigenvalue $\lambda=:x^2\geq 0$ of the positive Hermitian matrix $\chi:=M^\dagger M$.}\label{figsexticU}
\end{figure}

\begin{figure}
\begin{center}
\includegraphics[scale=0.5]{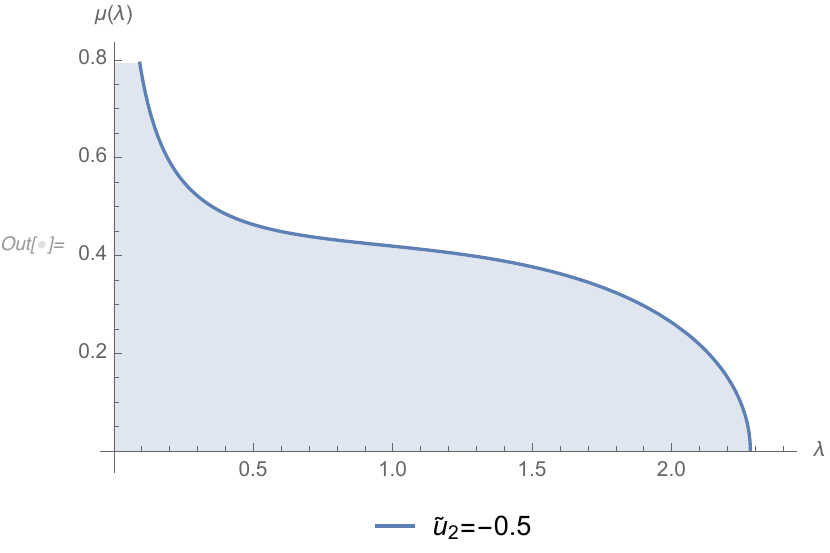}\quad \quad \includegraphics[scale=0.5]{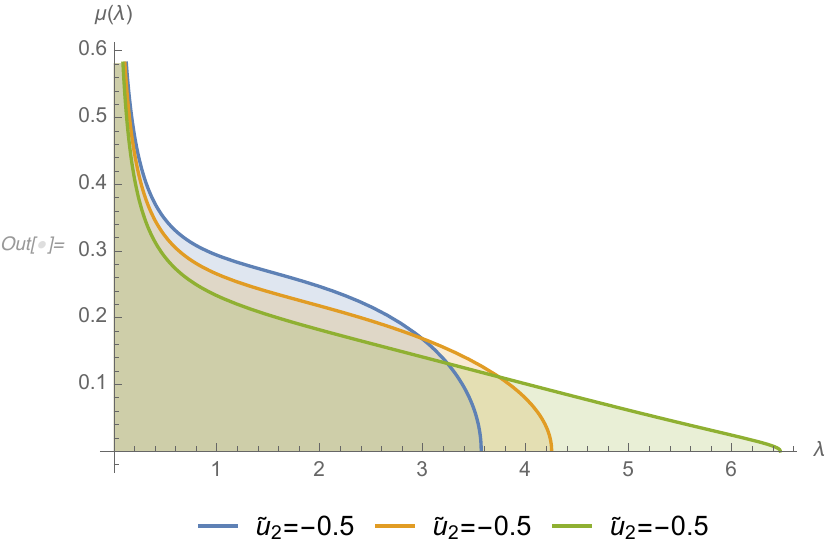}
\end{center}
\caption{Eigenvalue distributions along the axis $\tilde{u}_2$, at the fixed point $\text{FPB}$ (on left) and toward $\text{FPA}$ (on right).}\label{figsexticdis} 
\end{figure}

\begin{figure}
\begin{center}
\includegraphics[scale=0.8]{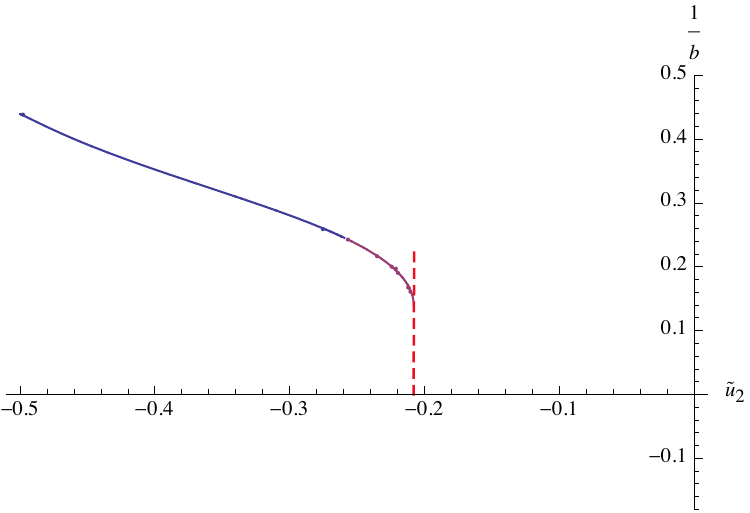}
\end{center}
\caption{The discontinuity for $b$ near $\tilde{u}_2 \approx -0.2$.}\label{figsexticdis2} 
\end{figure}

Local interactions of degrees higher than $4$ can be expressed as well in terms of the effective renormalized mass $\tilde{u}_2$. Indeed, we can compute the $\dot{\tilde{u}}_4^{(W)}$ from the explicit expression \eqref{explicitu4}, and if we compare it with the flow equation \eqref{flowequationu4tilde}, we deduce the expression of $\tilde{u}_6$, which can be used to deduce the expression for $\tilde{u}_8$ from its flow equation, and so one. We then deduce in principle the expressions for $\tilde{u}_{2n}$, even though concretely the expressions become rapidly intractable as $n>5$. In Figure \ref{figu6closed}, we show the behavior of couplings from $n=1$ to $n=5$, and the behavior of the effective potential along the $\tilde{u}_2$ axis, between fixed points $\text{FPA}$ and $\text{FPB}$ for a sextic truncation in Figure \ref{figsexticU}; In Figures \ref{figsexticdis}-\ref{figsexticdis2} we can identified the corresponding eigenvalues distributions due to the confining, computed using formula \eqref{formulamu} of appendix \ref{App6}. Starting from $\tilde{u}_2\approx -0.5$, and moving toward positive masses, we show that the width of the eigenvalue distribution increases until $\tilde{u}_2\approx -0.22$, where the distribution becomes unbounded. Indeed, up to this point, the potential becomes unbounded from below, as Figure \ref{figsexticU} shows, and eigenvalues are repelled from the origin. Hence, up to this point, there are no finite values for the parameter $b$ (see formula \eqref{formulamu}, where we make the assumption $a=0$).
Spontaneously, the value of $b^{-1}$ takes a finite value $b^{-1}\approx 0.15$ for $\tilde{u}_2\approx -0.22$, a discontinuity reminiscent of a first order phase transition. The results are reminiscent of the phase transition from region I to region II recalled in section \ref{seceqscheme2}, once again we discover a \textbf{discontinuous} (first order) phase transition with respect to the parameter $b$, between a condensed to a dilute phase for the eigenvalues.
\begin{claim}\label{claimWard1}
There exists a first-order phase transition between a high-temperature dilute phase and a condensed phase, with order parameter $b^{-1}$ jumping from $b^{-1}=0$ to $b^{-1}\approx 0.15$ for a temperature scale $\tilde{u}_2\approx -0.22$.  
\end{claim}
The values $(\tilde{u}_2,\tilde{u}_4,\tilde{u}_6)\approx (-0.22,0.77,-0.37)$ have to be compared with the corresponding values $(\tilde{u}_2,\tilde{u}_4)\approx (-0.22,0.92,-0.49)$ of the sextic truncation. 
\medskip

\begin{figure}
\begin{center}
\includegraphics[scale=0.8]{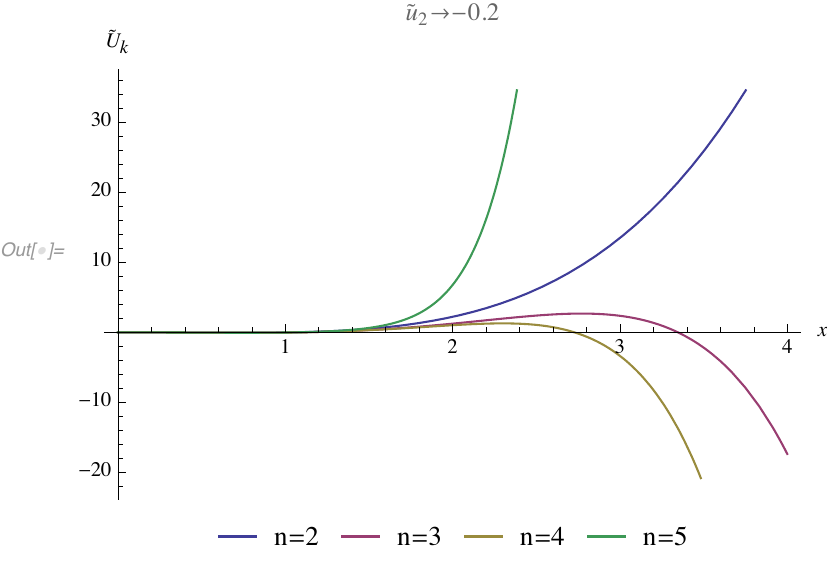}
\end{center}
\caption{Illustration of the poorly reliable shape of the potential for $\tilde{u}_2=-0.2$.}\label{fig33}
\end{figure}

\begin{figure}[h!]
\begin{center}
\includegraphics[scale=0.5]{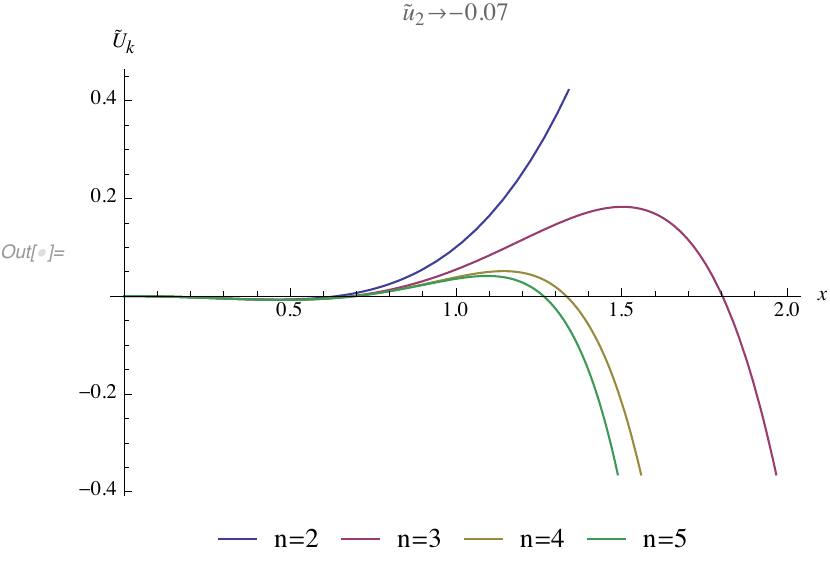}\qquad \includegraphics[scale=0.5]{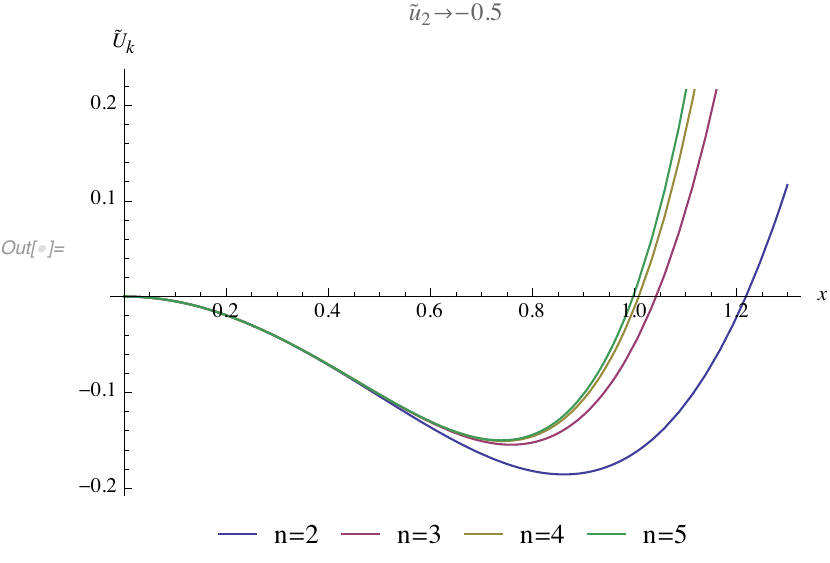}
\end{center}
\caption{The convergence of the potential in the vicinity of the two fixed points, for quartic (blue curve), sextic (purple curve), octic (yellow curve) and dectic (green curve).}\label{potentialconvergence}
\end{figure}

Now it becomes necessary to try to respond if the above results are reliable. This question will be the subject at the end of this subsection and of the next one. Note that on some regions along the mass axis, the sign of the larger order coupling changes from $n$ to $n+1$, and the shape of the effective potential is poorly reliable for some points on the interval between fixed-points -- see Figure \ref{fig33} for instance. It has however been noticed that the region where the larger coupling is positive increases with the degree of the truncation, and moves, from $\text{FPB}$, toward the positive mass region. Figure \ref{figu6closed} shows this explicitly, and it can be checked that this phenomenon is repeated for higher-order truncations. Furthermore, in some points along the $\tilde{u}_2$ axis, the rate of convergence is faster enough to make a reliable prediction. This is the case, for instance, for the potential near the fixed point $\text{FPB}$, as shown in Figure \ref{potentialconvergence} up to order $n=5$. Hence, we conclude that despite the instabilities observed at first orders of the vertex expansions, the results are reliable enough and make the prediction about first-order phase transition robust; higher-order couplings simply shift from right to the transition point, as Figure \ref{figsexticU} shows. In Figure \ref{distributionconvergence}, we can see the eigenvalue distribution for truncations up to order $5$ at $\text{FPB}$ and the dependency of the transition point are summarized in Table \ref{tablesummarise}. Note that the eigenvalue distribution around $\text{FPB}$ remains close to the sextic prediction. Finally, note that the quartic order predictions are very poorly reliable because they predict only a continuous phase transition between a low-temperature regime where eigenvalues are spread around a non-zero vacuum and a high-temperature regime where eigenvalues are spread around the zero vacuum. 

\medskip

We conclude this part by making the following important remark. The effective couplings $\tilde{u}_{2n} \tilde{\sigma}^{3n/2}$ have the power count $(3-n)/2$. Thus, by adding the couplings $\tilde{u}_{8}$ and $\tilde{u}_{10}$, we have included irrelevant effects, in competition with other effects of the same order (relative to the Gaussian counting), derivative couplings of higher order for example. Let us remark that it seems difficult under these conditions to have complete confidence in our estimates of high-order couplings, in any case, this does not constitute a sufficient argument to doubt the satisfaction \ref{claimWard1}, which moreover seems to agree with the predictions of vertex expansion. 
\pagebreak

\begin{multicols}{2}

\centering
$\vcenter{\hbox{\includegraphics[scale=0.5]{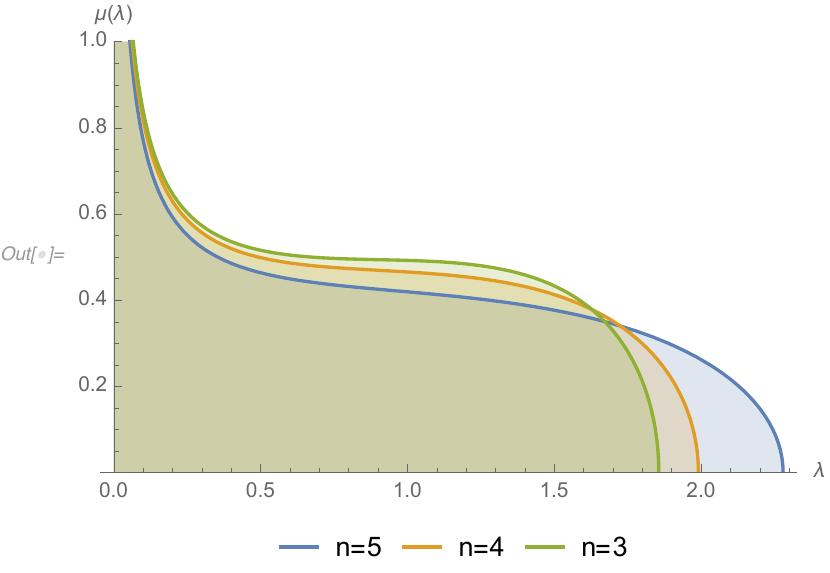}}}$
 \captionof{figure}{Eigenvalue distribution at $\text{FPB}$.}\label{distributionconvergence}


\columnbreak
\begin{center}
\begin{tabular}{|C{2.2cm}||C{1cm}|C{0.8cm}|C{1.2cm}|}
\hline  Truncation & $\tilde{u}_2$ &  $b_c$ & Order \\
\hline  n=2 & 0 & 4.06 & Second \\
\hline n=3 & -0.22 & 6.23 & First \\
\hline n=4 & -0.20 & 5.60 & First \\
\hline n=5 & -0.11 & 61 & First \\
\hline 
\end{tabular}
\captionof{table}{Summary of the properties of the transition point for truncations up to order $n=5$. As expected, the transition remains first order since $n>0$, and the transition point moves toward positive mass as $n$ increases.}\label{tablesummarise}
\end{center}
\end{multicols}
\subsubsection{Second modified Ward identity.} 
\begin{figure}
\begin{center}
\includegraphics[scale=0.55]{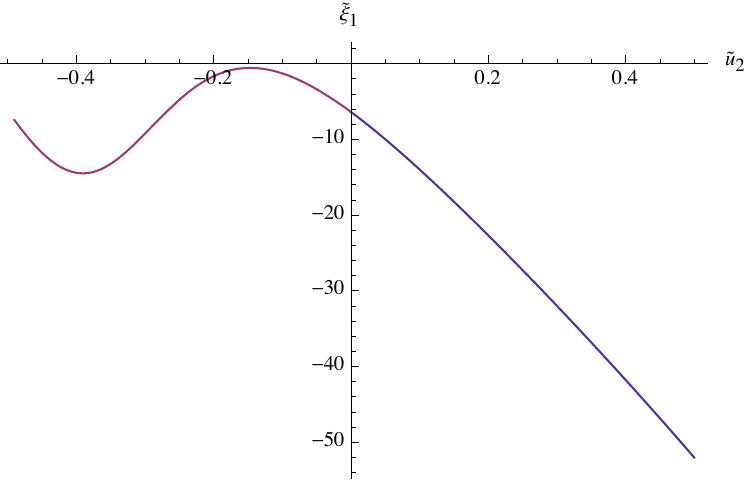}\qquad \includegraphics[scale=0.55]{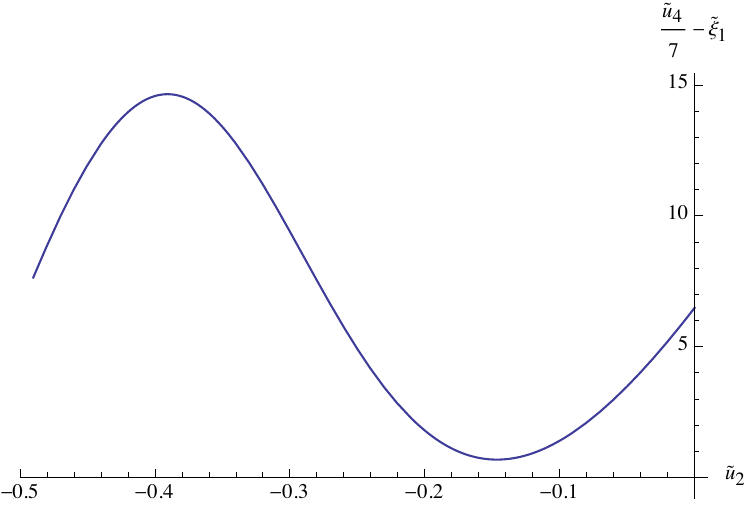}
\end{center}
\caption{On the left: The solution $\tilde{\xi}_1(\tilde{u}_2)$ rescaled with a factor $\tilde{\sigma}^{-3/2}$, in the negative and positive region. On the right: Comparison with the solution $\tilde{u}_4/7$.}\label{figu6W}
\end{figure}

In this section we then consider the sextic coupling $\tilde{u}_6$, which can be expressed using the second modified Ward identity \eqref{SmodifiedWI}. Setting $p_1^\prime \to p_1$, this identity reads:

\begin{align}
\nonumber \N\int \rho(q) \Bigg[1+&\frac{\extd R_k}{\extd p_1}(0,q)\Bigg] G_{k}^2(0,q)\bigg(\gamma_{k,q,0,0,0,0,0}^{(6)}-2(\gamma^{(4)}_{k,q,0,0,0})^2 G_{k}(0,q)\bigg)\\
&+\frac{\extd}{\extd p_1} \gamma_{k,p_1,0,0,0}^{(4)}\bigg\vert_{p_1=0}=0\,.\label{eqWard61TWO}
\end{align}
Furthermore, the expression of the derivative of $\gamma_k^{(4)}$ with respect to the external momenta can be obtained from the flow equation, because of the constraint $\tilde{\nu}=0$. Indeed, coming back on the computation of $\tilde{\nu}$, but taking into account the dependency of the effective vertex on the external momenta, we have:
\begin{align}
\nonumber k \frac{\extd Z}{\extd k}=&-\frac{1}{8} \left(\frac{\extd}{\extd p_1} \gamma_{k,p_1,0,0,0}^{(4)}+\frac{\extd}{\extd p_1} \gamma_{k,p_1,p_1,0,0}^{(4)}\Big\vert_{p_1=0}\right) \sum_{q} \frac{\dot{R}_k(q,0)}{(k+u_2)^2}\\
&-\frac{u_{4}}{8\N} \frac{\extd }{\extd p_1}\sum_{q} \frac{\dot{R}_k(q,p_1)}{(k+u_2)^2}\Big\vert_{p_1=0}\,,
\end{align}
or explicitly, because $\nu=0$:
\begin{equation}
0=\frac{u_{4}}{8\pi} \frac{k^{3/2}}{(k+u_2)^2}-\frac{7 k^{5/2}}{24\pi} \left(\frac{\extd}{\extd p_1} \gamma_{k,p_1,0,0,0}^{(4)}\Big\vert_{p_1=0}+\frac{\extd}{\extd p_1} \gamma_{k,p_1,p_1,0,0}^{(4)}\right)  \frac{1}{(k+u_2)^2}\,,
\end{equation}
leading to the remarkably simple result:
\begin{equation}
\boxed{k\frac{\extd}{\extd p_1} \gamma_{k,p_1,0,0,0}^{(4)}\Big\vert_{p_1=0}=\frac{1}{7}\frac{u_4}{\N}\,,}\label{solutionderiv}
\end{equation}
where we used the condition, coming from the chain rule:
\begin{equation}
k\frac{\extd}{\extd p_1} \gamma_{k,p_1,p_1,0,0}^{(4)}\Big\vert_{p_1=0}=2 k\frac{\extd}{\extd p_1} \gamma_{k,p_1,0,0,0}^{(4)}\Big\vert_{p_1=0}\,.\label{chainrule}
\end{equation}
Now, let us return to the computation of the Ward identity \eqref{eqWard61TWO}. As explained before, the terms proportional to $\partial_{p_1}R_k(0,q)$ can be computed using the local truncation, and we get:

\begin{align}
\nonumber &\N\int \rho(q) \Bigg[\frac{\extd R_k}{\extd p_1}(0,q)\Bigg] G_{k}^2(0,q)\bigg(\gamma_{k,q,0,0,0,0,0}^{(6)}-2(\gamma^{(4)}_{k,q,0,0,0})^2 G_{k}(0,q)\bigg)\\
&=-\frac{1}{\N}\frac{\left(\frac{1}{\tilde{\sigma }}\right)^{3/2}}{6\tilde{Z} k^2\pi}\frac{1}{(1+\tilde{u}_2)^2}\bigg(\frac{u_6}{3}-\frac{u_4^2}{4 \tilde{Z}k} \frac{1}{1+\tilde{u}_2}\bigg)\,.
\end{align}
The remaining terms can be computed using the same strategy as for the quartic case (see \eqref{methodQ4}). Then, we construct a contribution like $\sum_q G^2_k(0,q) \ddot{\Gamma}_k^{(2)}(q)$ from the first term (with $\gamma_k^{(6)}$), which is exactly compensated by the same contribution coming from the term with two $\gamma_k^{(4)}$. Explicitly we get:
\begin{align}
\nonumber &\int \rho(q) G_{k}^2(0,q)\bigg(\gamma_{k,q,0,0,0,0,0}^{(6)}-2(\gamma^{(4)}_{k,q,0,0,0})^2 G_{k}(0,q)\bigg)\\\nonumber 
&=-\frac{1}{\N^2\mathcal{L}_2^2} \, \frac{\extd^2}{\extd s^2} g(k,\tilde{u}_2)+\frac{2 u_4}{\N^2} \left(\frac{\mathcal{L}_3}{\mathcal{L}_2^2}\right)\frac{\extd}{\extd s} g(k,\tilde{u}_2)\\
&\quad +\frac{1}{\N^2\mathcal{L}_2^2}\frac{\extd}{\extd s} g(k,\tilde{u}_2) \frac{\extd}{\extd s} \mathcal{L}_2+\frac{1}{\N^2\mathcal{L}_2} \frac{\extd}{\extd s} \mathcal{L}_2\,,
\end{align}
and the Ward identity reads finally (remembering that $\tilde{Z}=1$):
\begin{align}
&\nonumber-\frac{1}{\mathcal{L}_2^2} \, \frac{\extd^2}{\extd s^2} g(k,\tilde{u}_2)+2 \tilde{u}_4 k^2\left(\frac{\mathcal{L}_3}{\mathcal{L}_2^2}\right)\frac{\extd}{\extd s} g(k,\tilde{u}_2)+\frac{1}{\mathcal{L}_2^2}\frac{\extd}{\extd s} g(k,\tilde{u}_2) \frac{\extd}{\extd s} \mathcal{L}_2\\
&+\frac{1}{\mathcal{L}_2} \frac{\extd}{\extd s} \mathcal{L}_2-\frac{\left(\frac{1}{\tilde{\sigma }}\right)^{3/2}}{6\pi}\frac{k}{(1+\tilde{u}_2)^2}\bigg(\frac{\tilde{u}_6}{3}-\frac{\tilde{u}_4^2}{4} \frac{1}{1+\tilde{u}_2}\bigg)+\frac{\extd}{\extd p_1} \gamma_{k,p_1,0,0,0}^{(4)}\bigg\vert_{p_1=0}=0\,.\label{secondWardeq}
\end{align}
This equation can be solved for $\frac{\extd}{\extd p_1} \gamma_{k,p_1,0,0,0}^{(4)}\vert_{p_1=0}=:\xi_1$, and we then deduce the second quantity $\frac{\extd}{\extd p_1} \gamma_{k,p_1,p_1,0,0}^{(4)}\vert_{p_1=0}=:\xi_2$ from equation \eqref{solutionderiv}. The behavior of $\xi_1$ and of the solution $\tilde{u}_4/7$ is shown in Figure \ref{figu6W}, and are again continuous at $\tilde{u}_2=0$. Obviously, the solutions do not match one with the other, and the Ward identity \eqref{secondWardeq} indeed fails at the order of approximation we considered. This point requires a deeper analysis which we reserve for a forthcoming work. We suspect that the computation of sub-leading interactions like $\xi_1$ could require going beyond the derivative expansion, including for instance higher order derivative interactions, like for instance $\sim Z_2(p_1^2+p_2^2)$ in the $2$-point function truncation (see \cite{canet2003nonperturbative}). In that way, the Ward identity \label{secondWardeq} will be used to fix the value of this interaction and so one; what could be improved also the computation of local interactions, for which we pointed out in the previous section the poverty of the predictions for high orders, which could be attributable to the fact of having ruled out the effect of derivative couplings. As mentioned before, we plan to address this question in the future. 

\subsection{Local approximation: Scheme 1}
\begin{figure}
\begin{center}
\includegraphics[scale=0.53]{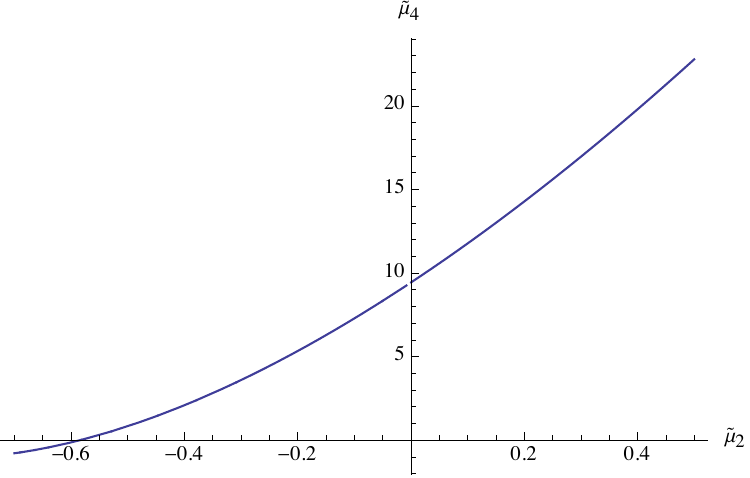}\qquad \quad\includegraphics[scale=0.53]{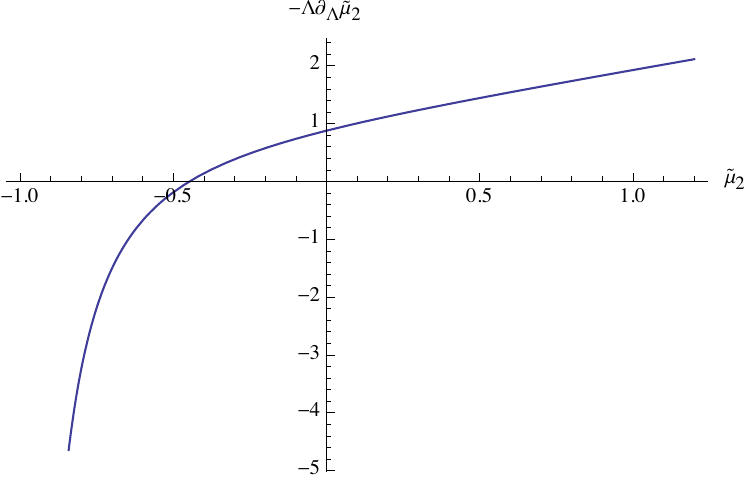}
\end{center}
\caption{On the left: Behavior of the solution for the effective quartic coupling. On the right: Behavior of the beta function for $\tilde{\mu}_2$ ($m=0$).}\label{figWardS2}
\end{figure}

Here, we consider the Ward identities for scheme 1. In this case, the variation $\delta E_k \equiv \delta E_\Lambda$ reads, in the limit $p_1^\prime \to p_1$:
\begin{align}
\nonumber \delta E_\Lambda(p_1^\prime,p_1,p_2)&= (p_1^\prime-p_1) \frac{\extd }{\extd p_1} (C^{(\Lambda)}_{p_1p_2})^{-1}\\
&= \frac{2(p_1^\prime-p_1)}{\chi_\Lambda(p_1)+\chi_\Lambda(p_2)} \left(1-\frac{p_1+p_2+m}{\chi_\Lambda(p_1)+\chi_\Lambda(p_2)}\chi^\prime_\Lambda(p_1)\right)\,,
\end{align}
and we have:
\begin{align}
\nonumber\sum_{p_2}\frac{\extd }{\extd p_1} (C^{(\Lambda)}_{p_1p_2})^{-1} (G^2_\Lambda)_{p_1,p_2}=&\frac{2(\chi_\Lambda(p_1)+\chi_\Lambda(p_2))}{(\Gamma_\Lambda^{(2)}(\chi_\Lambda(p_1)+\chi_\Lambda(p_2))+2(p_1+p_2+m))^2}\\
&\times \left(1-\frac{p_1+p_2+m}{\chi_\Lambda(p_1)+\chi_\Lambda(p_2)}\chi^\prime_\Lambda(p_1)\right)\,.
\end{align}
Due to the fact that $\chi_\Lambda(p_1)$ do not vanish  only on the interval $p_1 \in [\Lambda,4\sigma]$ (as $n\to \infty$), we must have $\chi^\prime_\Lambda(p_1)\simeq 0$ as soon as $\Lambda \neq 0$. Hence, the Ward identity \eqref{WardId3} reads in that case, with the momenta configuration \eqref{momentaconfig4pts}, but setting $p_1^\prime \to p_1 \to 0$:
\begin{equation}
\boxed{\sum_{p_2}\,\frac{2 \gamma_{\Lambda,p_2,0,0,0}^{(4)} \,\chi_\Lambda(p_2)}{(\Gamma_\Lambda^{(2)}(0,p_2)\chi_\Lambda(p_2)+2(p_2+m))^2}\,\,+\,\frac{\Gamma_\Lambda^{(2)}(\vec{p}\,^\prime)- \Gamma_\Lambda^{(2)}(\vec{p}\,)}{p_1^\prime-p_1}\Bigg\vert_{p_2=q_2} =0\,.}\label{WardId4S2}
\end{equation}
Because of the truncation \eqref{truncation}, we have:
\begin{equation}
\lim_{p_1^\prime\to p_1} \frac{\Gamma_\Lambda^{(2)}(\vec{p}\,^\prime)- \Gamma_\Lambda^{(2)}(\vec{p}\,)}{p_1^\prime-p_1}\Bigg\vert_{p_2=p_1=0}=2 z(\Lambda)\,.
\end{equation}
To compute the sum of the left-hand side of equation \eqref{WardId4S2}, we use the same strategy as in the previous section. 
Because of the flow equation \eqref{flowequ2S1}, and using the same arguments as before, we have, in the deep IR:
\begin{equation}
\frac{\partial}{\partial \Lambda} \Gamma_\Lambda^{(2)}(0,p_2)=-\gamma^{(4)}_{\Lambda,p_2,0,0,0} L_2(\Lambda)\,,
\end{equation}
where:
\begin{equation}
L_n(\Lambda):=2^{1-n}\,\Lambda^{-1}\rho(\Lambda) (1+z)^{-2} \frac{1+\tilde{m}}{(1+\tilde{\mu}_2)^n}\,,
\end{equation}
then:
\begin{align}
\sum_{p_2}\,\frac{2 \gamma_{\Lambda,p_2,0,0,0}^{(4)} \,\chi_\Lambda(p_2)}{(\Gamma_\Lambda^{(2)}(0,p_2)\,\chi_\Lambda(p_2)+2(p_2+m))^2}=-\frac{2/\N}{L_2(\Lambda)}\sum_{p_2}\,\frac{\left(\frac{\partial}{\partial \Lambda} \Gamma_\Lambda^{(2)}(0,p_2)\right) \,\chi_\Lambda(p_2)}{(\Gamma_\Lambda^{(2)}(0,p_2)\,\chi_\Lambda(p_2)+2(p_2+m))^2}\,,
\end{align}
which can be rewritten as:
\begin{align}
\nonumber\sum_{p_2}&\,\frac{\left(\frac{\partial}{\partial \Lambda} \Gamma_\Lambda^{(2)}(0,p_2)\right) \,\chi_\Lambda(p_2)}{(\Gamma_\Lambda^{(2)}(0,p_2)\,\chi_\Lambda(p_2)+2(p_2+m))^2}=-\frac{\partial}{\partial \Lambda} \underbrace{\sum_{p_2} \,\frac{1}{\Gamma_\Lambda^{(2)}(0,p_2)\,\chi_\Lambda(p_2)+2(p_2+m)}}_{:=\N A(m,\tilde{\mu}_2)}\\
&+\sum_{p_2} \frac{\Gamma_\Lambda^{(2)}(0,p_2) \,\delta_n(\Lambda-p_2)}{(\Gamma_\Lambda^{(2)}(0,p_2)\,\chi_\Lambda(p_2)+2(p_2+m))^2}\,.
\end{align}
The second term can be computed from the same strategy as we used for the computation of the loops $L_n(\Lambda)$, we get:
\begin{equation}
\sum_{p_2} \frac{\Gamma_\Lambda^{(2)}(0,p_2) \,\delta_n(\Lambda-p_2)}{(\Gamma_\Lambda^{(2)}(0,p_2)\,\chi_\Lambda(p_2)+2(p_2+m))^2}=\frac{1}{2} \frac{z(\Lambda+\bar{\mu}_2)\,\rho(\Lambda)}{(z(\Lambda+\bar{\mu}_2)+\Lambda+m)^2}\,,
\end{equation}
and, because of the definition \eqref{defdimensionless} for $\tilde{\mu}_2$:
\begin{equation}
\sum_{p_2} \frac{\Gamma_\Lambda^{(2)}(0,p_2) \,\delta_n(\Lambda-p_2)}{(\Gamma_\Lambda^{(2)}(0,p_2)\,\chi_\Lambda(p_2)+2(p_2+m))^2}=\frac{1}{2(1+z)^2\Lambda} \frac{(z+(1+z) \tilde{\mu}_2-\tilde{m})\,\rho(\Lambda)}{(1+\tilde{\mu}_2)^2}\,.
\end{equation}
The first term can be computed from the same observation as before: Only the vicinity of $p_2=\Lambda$ is relevant. Hence, in the deep IR, we make use of the truncation, in a region where it is assumed to have good convergence properties, and we get after a tedious calculation:
\begin{align}
&\nonumber \pi  (z+1)A(m,\tilde{\mu}_2)\approx\pi  \Lambda  \tilde{\mu }_2-\pi  \tilde{\mu }_2 \sqrt{\Lambda  \left(\frac{2}{\tilde{\mu }_2}+\Lambda \right)}+\pi\\
&+2 \sqrt{2} \left[\sqrt{\Lambda } \left(\sqrt{\tilde{\mu }_2} \cot ^{-1}\left(\sqrt{\tilde{\mu }_2}\right)+z\right)-\sqrt{m} (1+z) \tan ^{-1}\left(\sqrt{\frac{\lambda}{m }}\right)\right]\,,
\end{align}
which can be expanded in power of $\Lambda$, and keeping only the leading orders, we have:
\begin{align}
A(m,\tilde{\mu}_2)\approx \frac{1}{1+z}+\frac{\sqrt{2} \left(-\pi\sqrt{\tilde{\mu }_2}+2 \sqrt{\tilde{\mu }_2} \cot ^{-1}\left(\sqrt{\tilde{\mu }_2}\right)-2\right)}{\pi  (1+z)}\sqrt{\Lambda}+\mathcal{O}(\Lambda)\,,
\end{align}
then:
\begin{align}
\nonumber-\Lambda \frac{\partial}{\partial \Lambda}\, A(m,\tilde{\mu}_2)=&\frac{\eta}{1+z}-\frac{\sqrt{2} \left(-\pi\sqrt{\tilde{\mu }_2}+2 \sqrt{\tilde{\mu }_2} \cot ^{-1}\left(\sqrt{\tilde{\mu }_2}\right)-2\right)}{2\pi  (1+z)}\sqrt{\Lambda}\\
& -\frac{\sqrt{2} \left(-\frac{1}{2} \pi  \sqrt{\frac{1}{\tilde{\mu }_2}}-\frac{1}{\tilde{\mu }_2+1}+\frac{\cot ^{-1}\left(\sqrt{\tilde{\mu }_2}\right)}{\sqrt{\tilde{\mu }_2}}\right)}{\pi  (1+z)} \sqrt{\Lambda} \,\dot{\tilde{\mu}}_2\,.
\end{align}
Finally, because of the asymptotic expression for $\Lambda \ll 1$:
$
\rho(\Lambda) \sim \frac{2 \sqrt{2} \sqrt{\Lambda}}{\pi }\,,
$
the leading order Ward identity (of order $\Lambda^{-1/2}$) reads finally, for $m=0$ as:
\begin{align}
\sqrt{2}\pi (1+\tilde{\mu}_2)^2(1+z) \eta =0\,,
\end{align}
which implies that 
\begin{equation}
\boxed{\eta=0\,,}
\end{equation}
and then we can fix the asymptotic value for $z$, we choose: $z=1$, and the next to leading order term of Ward identity (of order $\mathcal{O}(1)$) is:
\begin{align}
\nonumber 1+ 4(1+&\tilde{\mu}_2)^2 \Bigg[\,\frac{1}{2}\left(-\pi \sqrt{\tilde{\mu }_2}+2 \sqrt{\tilde{\mu }_2} \cot ^{-1}\left(\sqrt{\tilde{\mu }_2}\right)-2\right)\\
&+\left(-\frac{1}{2} \pi  \sqrt{\frac{1}{\tilde{\mu }_2}}-\frac{1}{\tilde{\mu }_2+1}+\frac{\cot ^{-1}\left(\sqrt{\tilde{\mu }_2}\right)}{\sqrt{\tilde{\mu }_2}}\right)  \,\dot{\tilde{\mu}}_2\,\Bigg] -4(1+2\tilde{\mu}_2)=0\,,
\end{align}
which can be solved for $\dot{\tilde{\mu}}_2$ as:
\begin{equation}
\boxed{\Lambda\partial_\Lambda{\tilde{\mu}}_2=\frac{\frac{1}{2}\left(\pi\sqrt{\tilde{\mu }_2}-2 \sqrt{\tilde{\mu }_2} \cot ^{-1}\left(\sqrt{\tilde{\mu }_2}\right)+2\right)+\frac{4(1+2\tilde{\mu}_2)-1}{4(1+\tilde{\mu}_2)^2}}{\frac{\cot ^{-1}\left(\sqrt{\tilde{\mu }_2}\right)}{\sqrt{\tilde{\mu }_2}}-\frac{1}{2} \pi  \sqrt{\frac{1}{\tilde{\mu }_2}}-\frac{1}{\tilde{\mu }_2+1}}\,.}
\end{equation}
Comparing with the flow equation \eqref{flowequationmu2}, we deduce an explicit expression for $\tilde{\mu}_4$:
\begin{equation}
\boxed{\tilde{\mu}_{4}^{(W)}=\frac{12 \pi  \sqrt{\tilde{\mu }_2} \left(\tilde{\mu }_2+1\right) \left(12 \tilde{\mu }_2+7\right)}{7 \left(\sqrt{\tilde{\mu }_2} \left(\pi  \sqrt{\frac{1}{\tilde{\mu }_2}} \left(\tilde{\mu }_2+1\right)+2\right)-2 \left(\tilde{\mu }_2+1\right) \cot ^{-1}\left(\sqrt{\tilde{\mu }_2}\right)\right)}\,.}
\end{equation}
On Figure \ref{figWardS2}, we summarized the behavior of $\tilde{\mu}_{4}^{(W)}$ and $\Lambda\partial_\Lambda{\tilde{\mu}}_2$. The beta function has one fixed point, for the values:
\begin{equation}
\tilde{\mu}_2^{(*)}\approx -0.45\,,\qquad \tilde{\mu}_4^{(*)}\approx 1.47\,,
\end{equation}
with critical exponents:
\begin{equation}
\theta_* \approx +3.21\,.
\end{equation}

\begin{figure}
\begin{center}
\includegraphics[scale=0.53]{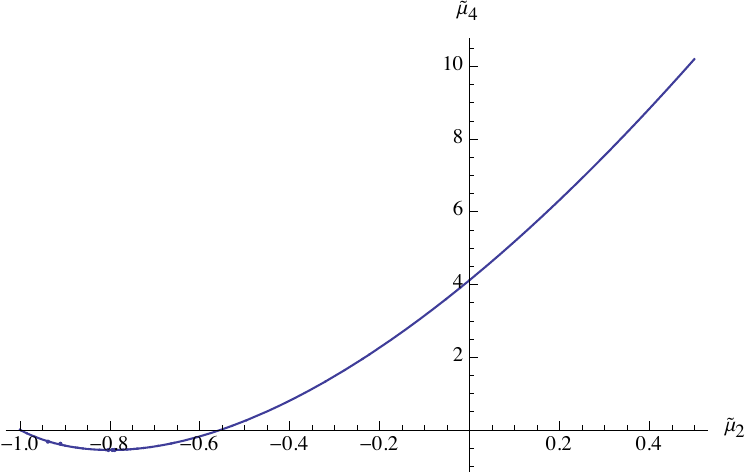}\qquad \quad\includegraphics[scale=0.53]{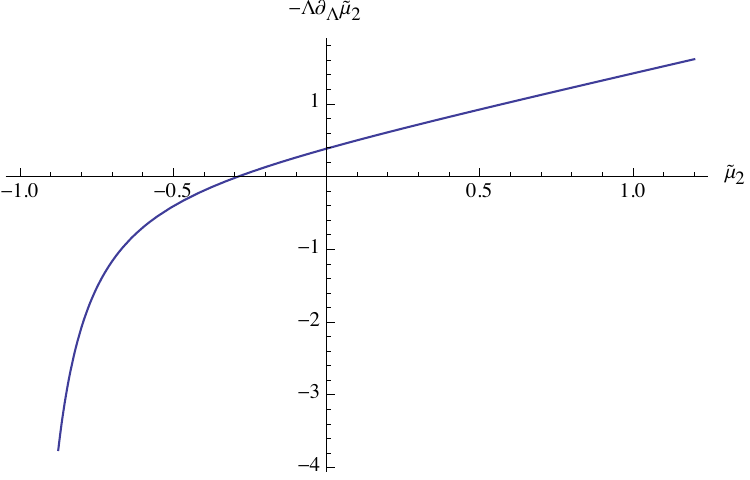}
\end{center}
\caption{On the left: Behavior of the solution for the effective quartic coupling (rescaled by a factor $(1+\tilde{m})^{-1})$. On the right: Behavior of the beta function for $\tilde{\mu}_2$ ($m\neq 0$).}\label{figWardS2m}
\end{figure}

For $m\neq 0$, the Ward identities can be investigated in the deep IR again, assuming that $\vert \tilde{m} \vert \gg 1$. At the leading order, we get, again $\eta=0$, but the value for $z$ is fixed by the term of order $\mathcal{O}(\tilde{m}^0,\Lambda^0)$, which reads:
\begin{equation}
2z+2(1+z)^2=0 \quad \Rightarrow \quad z_{\pm}=\frac{-3\pm \sqrt{5}}{2}\,.
\end{equation}
The positive solution $z_+\simeq -0.38$ ensures that $1+z_+\approx 0.62 >0$, and we retain only that one solution. At the order $\mathcal{O}(\tilde{m}^{-1})$, we have:
\begin{align}
\nonumber (1+&\tilde{\mu}_2)^2 \Bigg[\,\frac{1}{2}\left(-\pi \sqrt{\tilde{\mu }_2}+2 \sqrt{\tilde{\mu }_2} \cot ^{-1}\left(\sqrt{\tilde{\mu }_2}\right)-2\right)\\
&+\left(-\frac{1}{2} \pi  \sqrt{\frac{1}{\tilde{\mu }_2}}-\frac{1}{\tilde{\mu }_2+1}+\frac{\cot ^{-1}\left(\sqrt{\tilde{\mu }_2}\right)}{\sqrt{\tilde{\mu }_2}}\right)  \,\dot{\tilde{\mu}}_2\,\Bigg] -(z_++(1+z_+)\tilde{\mu}_2)(1+z_+)=0\,,
\end{align}
then:
\begin{equation}
\boxed{\Lambda\partial_\Lambda{\tilde{\mu}}_2\vert_{m\neq 0}=\frac{\frac{1}{2}\left(\pi \sqrt{\tilde{\mu }_2}-2 \sqrt{\tilde{\mu }_2} \cot ^{-1}\left(\sqrt{\tilde{\mu }_2}\right)+2\right)+\frac{(4-2\sqrt{5}+(3-\sqrt{5})\tilde{\mu}_2)}{2(1+\tilde{\mu}_2)^2}}{\frac{\cot ^{-1}\left(\sqrt{\tilde{\mu }_2}\right)}{\sqrt{\tilde{\mu }_2}}-\frac{1}{2} \pi  \sqrt{\frac{1}{\tilde{\mu }_2}}-\frac{1}{\tilde{\mu}_2+1}}\,.}
\end{equation}
The beta function vanishes for the value
\begin{equation}
\tilde{\mu}_2^{(*,m)}\approx -0.287\,,\qquad \tilde{\mu}_4^{(*,m)}\approx 1.57\,,
\end{equation}
which corresponds to the critical exponents:
\begin{equation}
\theta_*^{(m)}\simeq 1.55\,.
\end{equation}
Solving the flow equation for $\tilde{\mu}_2$, we get the expression for $\tilde{\mu}_4^{(W,m)}$:
\begin{equation}
\boxed{\tilde{\mu}_4^{(W,m)}=-\frac{24 \pi  \sqrt{\tilde{\mu }_2} \left(\tilde{\mu }_2+1\right) \left(\left(\sqrt{5}-5\right) \tilde{\mu }_2+2 \left(\sqrt{5}-3\right)\right)}{7 (1+\tilde{m}) \left(\sqrt{\tilde{\mu }_2} \left(\pi  \sqrt{\frac{1}{\tilde{\mu }_2}} \left(\tilde{\mu }_2+1\right)+2\right)-2 \left(\tilde{\mu }_2+1\right) \cot ^{-1}\left(\sqrt{\tilde{\mu }_2}\right)\right)}\,.}
\end{equation}
The Figure \ref{figWardS2m} summarizes the results for $m\neq 0$. 

\subsection{Short comment about Hermitian matrices}

Let us discuss briefly our construction for Hermitian matrices. The stochastic quantization of Hermitian matrices uses the same kind of equation as \eqref{eq1}, but includes the constraint:
\begin{equation}
M=M^\dagger\,,
\end{equation}
along the dynamics. In Appendix \ref{App1} we provided some details about the derivation of the corresponding Fokker-Planck equation, and as expected, the equilibrium theory is described by the quenched path integral:
\begin{equation}
Z_{\text{eq}}[L]:=\int \extd M \, e^{-2 \frac{H[M]}{T}+ \overline{L}\cdot M}\,,
\end{equation}
where here $\extd M$ is the standard measure over $\N \times \N$ Hermitian matrices manifold, see Appendix \ref{App1}. In the large $\N$ limit, and working in the eigenbasis of the disordered matrix, the equilibrium partition function becomes (setting again $T=2$):
\begin{equation}
Z_{\text{eq}}[L]\underset{\N\to \infty}{\longrightarrow}\int \extd M \, e^{- H_\infty[M]+ \overline{L}\cdot M}\,,\label{parteqcomplex2}
\end{equation}
where explicitly
\begin{equation}
H_\infty[M]:= \frac{1}{2}\sum_{\lambda_1,\lambda_2} \overline{M}_{\lambda_1\lambda_2} (\lambda_1+\lambda_2+a_1){M}_{\lambda_1\lambda_2}+\sum_{p=2}^\infty\,\frac{a_p \N^{-p+1}}{2p} \,\int_{-\infty}^{+\infty} \extd t \,\Tr \, (M(t))^{2p}\,.\label{HamiltonianCdef}
\end{equation}
Note that in this expression, and in contrast with the complex case, $\lambda_1$ and $\lambda_2$ are eigenvalues of the \textit{same} Wigner hermitian matrix. We focus on scheme 2 for this section, and in replacement with the local truncation \eqref{truncationPhi2}, we have:
\begin{equation}
\Gamma_k [M]= \frac{Z(k)}{2} \sum_{p_1,p_2} \overline{\Phi}_{p_1p_2}(p_1+p_2+2u_2(k)) \Phi_{p_1p_2}+ \sum_{n>1} \frac{u_{2n}(k)}{(2n) \N^{n-1}} \Tr (\Phi)^{2n}\,.\label{truncationPhi2H}
\end{equation}
Moreover, we keep the Litim regulator \ref{eqR}, but we add a factor $1/2$ in front of the definition of $\Delta S_k$. With this parametrization of the theory space, the $\beta$-functions can be easily computed, and we get for instance, for the $2$-points function (see \cite{lahoche20241} for more details):
\begin{equation}
\dot{\tilde{u}}_2=-(1+\tilde{\nu})\tilde{u}_2-\frac{\tilde{u}_{4}}{8\pi} \frac{J_{\tilde{\nu}}(k)}{(1+\tilde{u}_2)^2}\,,\label{equationflowu2S2HER}
\end{equation}
the definition of the dimensionless couplings being the same as for complex matrices. The equation is the same as for a complex matrix because of the normalization of the vertices. 
\medskip

The Ward identities can be deduced as for the complex case, by requiring the invariance of the partition function under some unitary transformation $M\to U^\dagger M U$, where $U\in \mathcal{U}(\N)$. Considering some infinitesimal transformation $U=I+i\epsilon$, with $\epsilon=\epsilon^\dagger$, the matrix field transforms as:
\begin{equation}
M\to M + i [M,\epsilon]+ \mathcal{O}(\epsilon^2)\,.
\end{equation}
Let us compute the variation of the kinetic action. We get, at the first order in $\epsilon$:
\begin{align}
\nonumber 2\delta_\epsilon S_{\text{kin}}&=i\sum_{p_1,p_2}  [M,\epsilon]_{p_2p_1} E_k(p_1,p_2){M}_{p_1p_2}+i\sum_{p_1,p_2} {M}_{p_2p_1} E_k(p_1,p_2)[M,\epsilon]_{p_1p_2}\\
&= i \sum_{p_1,p_1^\prime,p_2}\delta E_k(p_1,p_1^\prime,p_2) M_{p_2p_1}M_{p_1^\prime p_2} \epsilon_{p_1p_1^\prime}\,.
\end{align}
For the source term, we get in the same way:
\begin{align}
\delta_\epsilon S_{\text{source}}&= i \sum L_{p_1p_2} [M,\epsilon]_{p_1,p_2}\\
&=i \sum_{p_1,p_1^\prime,p_2} \left(L_{p_1p_2} M_{p_1^\prime p_2}-L_{p_2p_1^\prime} M_{p_2p_1}\right)\epsilon_{p_1p_1^\prime}\,.
\end{align}
Hence, the first order Ward identity reads finally:

\begin{proposition}
The partition function obeys the following differential equation:
\begin{align}
\sum_{p_2} \Bigg( \delta E_k(p_1,p_1^\prime,p_2) \frac{\partial^2}{\partial L_{p_2p_1}\partial {L}_{p_1^\prime p_2}} + {L}_{p_1p_2} \frac{\partial}{\partial {L}_{p_1^\prime p_2}}-{L}_{p_2 p_1^\prime } \frac{\partial}{\partial {L}_{p_2 p_1}}   \Bigg)Z_{\text{eq}}[L]=0\,.\label{WardId}
\end{align}
\end{proposition}
Using the same notations as in section \ref{planarWI}, we get:
\begin{align}
\sum_{p_2} \Bigg(\delta E_k(p_1,p_1^\prime,p_2) \left(G^{(2)}_{k,\vec{p}\,^T\vec{p}\,^\prime}+{\Phi}_{\vec{p}\,^T} \Phi_{\vec{p}\,^\prime}\right) + {L}_{\vec{p}}\, {\Phi}_{\vec{p}\,^\prime}-{L}_{\vec{p}\,^{\prime T}}\, {\Phi}_{\vec{p}\,^T}   \Bigg)=0\,,\label{WardId2}
\end{align}
where $\vec{p}:=(p_1,p_2)$ and $\vec{p}\,^T:=(p_2,p_1)$. Taking the second derivative with respect to the classical field $\Phi$, $\partial^2/\partial \Phi_{\vec{q}}\,\partial \Phi_{\vec{q}\,^\prime}$, and imposing the symmetric phase conditions:
\begin{align}
\nonumber \sum_{p_2} &-\delta E_k(p_1,p_1^\prime,p_2) G^{(2)}_{k,\vec{p}\,^T\,\vec{r}}\,\Gamma_{k,\vec{q}\,\vec{q}\,^\prime\,\vec{r}\,\vec{s}}^{(4)}\,G^{(2)}_{k,\vec{s}\,\vec{p}\,^\prime} + \sum_{p_2}  \Big[\delta E_k(p_1,p_1^\prime,p_2) \big(\delta_{\vec{p}\,^T\,\vec{q}}\,\delta_{\vec{p}\,^\prime\,\vec{q}\,^\prime}\\\nonumber
&+\delta_{\vec{p}\,^T\,\vec{q}\,^\prime}\,\delta_{\vec{p}\,^\prime\,\vec{q}\,}\big)+\left(\Gamma_k^{(2)}(\vec{p}\,)+R_k(\vec{p}\,)\right)\big(\delta_{\vec{p}\,\vec{q}}\,\delta_{\vec{p}\,^\prime\vec{q}\,^\prime}+\delta_{\vec{p}\,\vec{q}\,^\prime}\,\delta_{\vec{p}\,^\prime\vec{q}\,}\big)\\
&-\left(\Gamma_k^{(2)}(\vec{p}\,^{\prime T}\,)+R_k(\vec{p}\,^{\prime T}\,)\right)\Big(\delta_{\vec{p}\,^T\, \vec{q}}\,\delta_{\vec{p}\,^{\prime T}\vec{q}\,^\prime}+\delta_{\vec{p}\,^T\, \vec{q}\,^\prime}\,\delta_{\vec{p}\,^{\prime T}\vec{q}\,}\big)\Big]=0\,.\label{WardId3H}
\end{align}
This equation can be investigated as before, and the results are very similar. In particular, we get again that anomalous dimension have to vanish for the asymptotic flow in the deep IR regime. This condition has been found moreover in the tensorial field theory context, as a requirement for fixed point condition \cite{Lahoche_2019bb,Lahoche_2020b}.

\section{Conclusion and open issues}\label{concluding}

In this paper, we have presented a method going beyond the vertex expansion and exploiting Ward identities arising consequence of the symmetry breaking of the effective kinetics of the model. This second method imposes strong constraints on the renormalization group, notably imposing the cancellation of the anomalous dimension asymptotically towards the IR. For equilibrium theory this method confirms some predictions
of vertex expansion, but not all. In particular, we enforce the evidence in favor of the existence of a first-order phase transition, resulting in the confinement of the eigenvalues seems to be inevitable. 
\medskip

However, this conclusion remains partial, and the proposed formalism must still be completed by subsequent work. In particular, the role of higher-order derivative interactions must be carefully evaluated and seems to play a determining role when we involve interactions of a higher order than 6. This open issue will be the subject of future work. Moreover, some aspects have been ignored in this formalism. In particular, the behavior of the flow in the IR but not deep IR, where the specific shape of the momenta distribution could play a relevant role, has been ignored. Furthermore, the role of $1/\N$ corrections to the Wigner law, which could become significant in the deep IR as $k,\Lambda\sim 1/\N$, especially for sextic interactions, have been ignored as well and will be also the subject of forthcoming work. Finally, we only focused on first-order Ward identities, and it has been shown recently that next-to-leading-order Ward identities could be required for that kind of theory \cite{kpera2023anomalous}. 

\pagebreak

\appendix
\begin{center}
\begin{LARGE}
\textbf{Additional material}
\end{LARGE}
\end{center}

\section{Proof of lemma \ref{lemma1} (sketched)}\label{App7}

\begin{figure}[H]
\begin{center}
\includegraphics[scale=1]{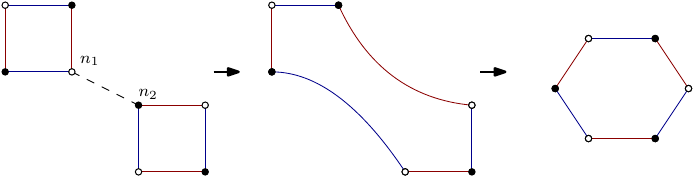}
\end{center}
\caption{Sum of two quartic vertices.}\label{figsum}
\end{figure}

In this Appendix, we provide a proof of the statement of the lemma \ref{lemma1}. We get the proof for complex matrices, but it can be easily generalized for Hermitian matrices. More precisely, we will prove the more general statement:
\begin{lemma}\label{lemma2}
Any 1PI Feynman diagram is contractible, in the sense that all the internal edges can be contracted and the resulting contraction is a connected vertex. 
\end{lemma}
This statement implies lemma \ref{lemma1}. 
We first define the sum of two connected vertices:
\begin{definition}
The sum of two vertices $v_1$ and $v_2$ is the vertex $v_3$ constructed as follow:
\begin{enumerate}
    \item We link together two nodes $n_1\in v_1$ and $n_2\in v_2$ of different colors with a dotted edge.
    \item We contract the dotted edge $(n_1n_2)$.
\end{enumerate}
Figure \ref{figsum} provides an example of the sum of two quartic vertices. 
\end{definition}

\begin{figure}[H]
\begin{center}
\includegraphics[scale=0.8]{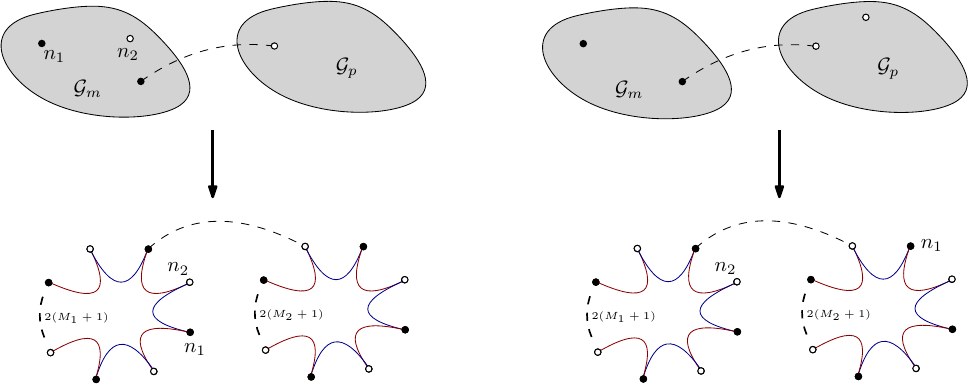}
\end{center}
\caption{The two allowed configuration when $\mathcal{G}_n^\prime$ is PI. The components $\mathcal{G}_m$ and $\mathcal{G}_p$ are 1PI. Only the configuration on the right is allowed by the recursion assumption.}\label{figallowedPI}
\end{figure}

Hence, any Feynman diagram involving different size vertices can be transformed as an equivalent Feynman diagram involving only quartic vertices, and we restrict our proof to the quartic sector. It is easy to check that the statement holds for 1PI Feynman diagrams involving two vertices. Now, we assume the statement to be true for 1PI Feynman diagrams involving $n$ Feynman diagrams, and we have to prove that it is true also at order $n+1$. Two moves allow us to make this. First, we have \textit{tadpole insertion}, which inserts a tadpole along some internal line:
\begin{equation}
\vcenter{\hbox{\includegraphics[scale=0.7]{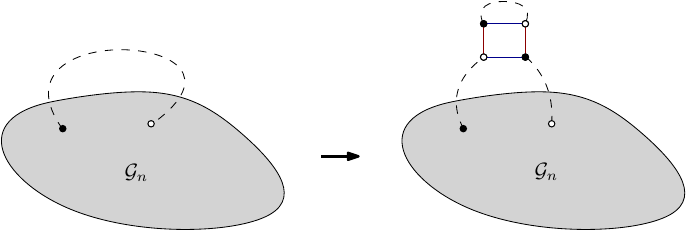}}}
\end{equation}
where the gray region materialize some 1PI Feynman diagram $\mathcal{G}_n$, of order $n$ with $2M$ external edges. We denote as $\ell$ the internal edge on which we performed the tadpole insertion; then if we cancel the edge $\ell$, we obtain a Feynman graph $\mathcal{G}_n^\prime$ with $2(M+1)$ external points:
\begin{equation}
\vcenter{\hbox{\includegraphics[scale=0.7]{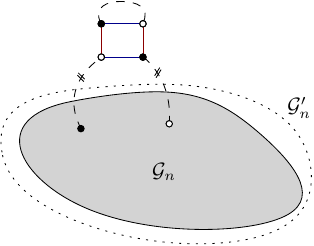}}}
\end{equation}
But because of the recursion hypothesis, the graph $\mathcal{G}_n^\prime$ has to be contractible; more precisely, we have to distinguish two cases. In the first one, $\mathcal{G}_n^\prime$ is 1PI itself, and then from the recursion hypothesis, the diagram $\mathcal{G}_n^\prime$ has to be contractible and the result of the contraction of all the internal edges is a connected vertex:
\begin{equation}
\vcenter{\hbox{\includegraphics[scale=0.7]{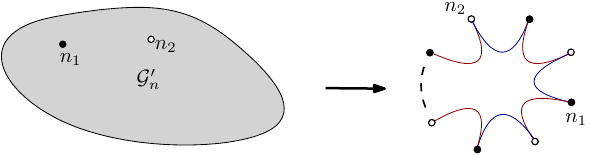}}}
\end{equation}
The diagram $\mathcal{G}_{n+1}$ will be planar (i.e. leading order) only if the two nodes $n_1$ and $n_2$ are contiguous. Hence the contractions off all the edges of $\mathcal{G}_n^\prime$ has to be of the form:
\begin{equation}
\vcenter{\hbox{\includegraphics[scale=0.9]{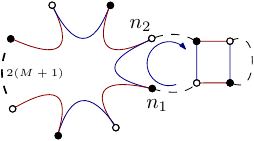}}}\,.
\end{equation}
Then, the remaining edges can be easily contracted:
\begin{equation}
\vcenter{\hbox{\includegraphics[scale=0.7]{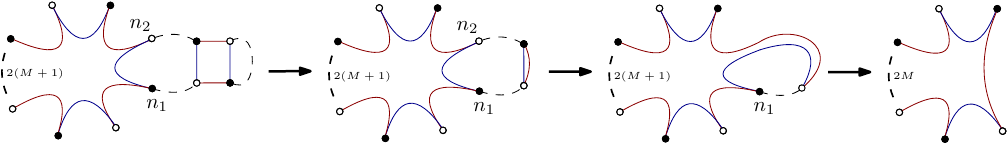}}}
\end{equation}
and the result agrees with the statement \ref{lemma2}. Now, let us investigate the case where $\mathcal{G}_n^\prime$ is not 1PI. All the allowed configurations are pictured in Figure \ref{figallowedPI}, and it is easy to check that, before contracting the remaining edge, the two nodes $n_1$ and $n_2$ have not to be contiguous because of the leading order assumption (as on the right of Figure \ref{figallowedPI}), and contracting the remaining edges, we find that the statement holds. 
\medskip

\begin{figure}[H]
\begin{center}
\includegraphics[scale=0.7]{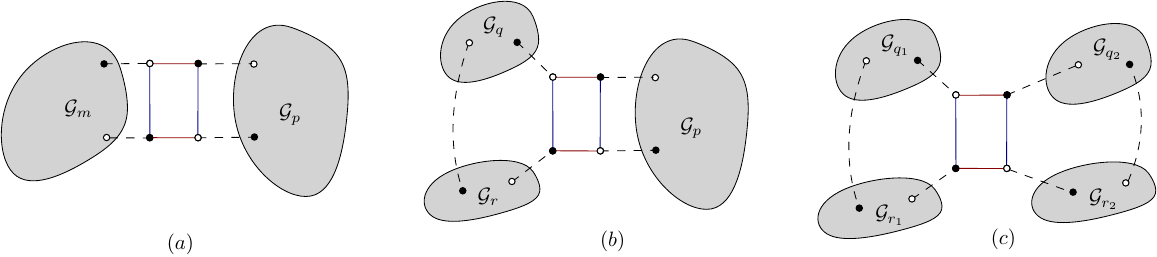}
\end{center}
\caption{All the 1PI decompositions for the second type insertions: all the gray blobs denote 1PI components.}\label{fig1PIdec2}
\end{figure}

A direct inspection shows that the other way to add a vertex is:
\begin{equation}
\vcenter{\hbox{\includegraphics[scale=0.7]{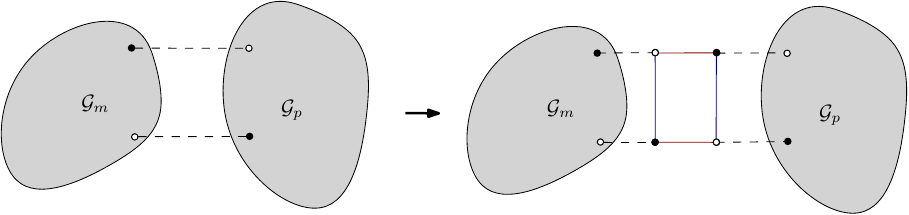}}}
\end{equation}
We have to investigate the different 1PI decomposition for the subgraphs $\mathcal{G}_m$ and $\mathcal{G}_p$. There are three 1PI decompositions, all pictured in Figure \ref{fig1PIdec2}. All the possibilities can be investigated one by one as for the previous case. Let us consider for instance the third case (c); all the 1PI components have to be contractible because of the recursion assumption. Furthermore, it is easy to check that the condition optimizing the number of created faces is the following:
\begin{equation}
\vcenter{\hbox{\includegraphics[scale=0.7]{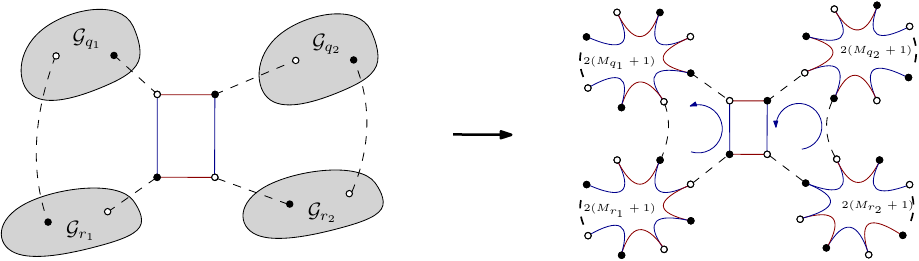}}}\,.
\end{equation}
Note that, because each vertex has weight $1/\N$, the number of faces has been increased by $1$
 exactly. Contracting the remaining edges, we get a connected vertex, and the statement holds again. 

\section{Computing eigenvalue spectra of $M^\dagger M$.}\label{App6}

\begin{figure}[H]
\begin{center}
\includegraphics[scale=1]{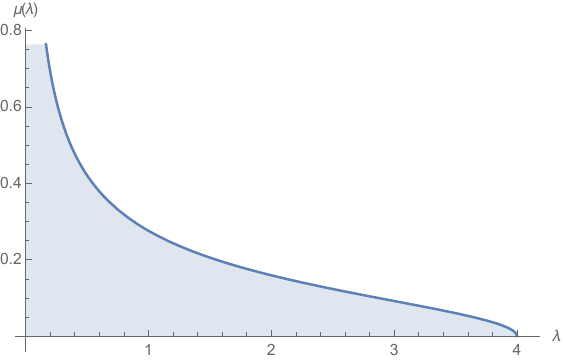}
\end{center}
\caption{Typical shape of the eigenvalue distribution with $\mu_1=1$.}\label{FigMP}
\end{figure}

In this section, we shortly review the solution of the problem to determine the eigenvalue spectra of $\chi=M^\dagger M$, in the large $\N$ limit, from the knowledge of the distribution law for the complex matrix $M$. We assume the partition is of the form:
\begin{equation}
Z:=\int \extd M \extd \bar{M} \, e^{-\mu_1 \Tr M^\dagger M - \Tr\, V(M,M^\dagger)}\,,
\end{equation}
where $V$ expands in power of $\chi$:
\begin{equation}
V(M,M^\dagger)=\sum_{p=1}^K \frac{\mu_p}{(p!)^2 \N^{p-1}}\, (M^\dagger M)^{p}\,.
\end{equation}
To perform the change of variable $\chi=M^\dagger M$ in the path integral, we use the clever expression of the identity:
\begin{equation}
1\equiv \int \extd \chi \,\delta\left(\chi-M^\dagger M\right)\,,
\end{equation}
where the integral is performed on the space of Hermitian matrices. The Dirac delta function can be furthermore rewritten in the Fourier representation for the $N(N+1)$ independent components of $\chi$:
\begin{equation}
\delta(\chi-M^\dagger M) \propto \int \extd \lambda \, e^{i\, \Tr \lambda (\chi-M^\dagger M)}\,,
\end{equation}
where the integration is performed on the hermitian matrix $\lambda$\footnote{It is suitable to add a small imaginary part proportional to the identity to make the integral well defined.}; furthermore, it is suitable to exploit the unitary invariance of the integral over $M$ to choose $\chi$ diagonal. Finally, the partition function reads: 
\begin{equation}
Z=\int \extd M \extd \bar{M} \extd \chi \extd \lambda \, e^{i\, \Tr \lambda (\chi-M^\dagger M)} e^{-\mu_1 \Tr \chi - \Tr\, V(\chi)}\,.
\end{equation}
The Gaussian integral over $M$ can be easily performed, 
\begin{equation}
\int \extd M \extd \bar{M}e^{-i\, \Tr \lambda M^\dagger M} \propto (\det \lambda)^{-N}\equiv  e^{-\N\, \Tr\,\ln (\lambda)}\,,
\end{equation}
such that the partition function becomes,
\begin{equation}
Z=\int \extd \chi \extd \lambda \,e^{-\N\, \Tr\,\ln(\lambda)-\mu_1 \Tr \chi - \Tr\, V(\chi)} e^{i\, \Tr\, \lambda \chi}\,.
\end{equation}
Let us make the change of variable $\lambda \to \sigma:= \chi^{\frac{1}{2}} \lambda \chi^{\frac{1}{2}}$, the Jacobian of the transformation is:
\begin{align}
\prod_i \extd\lambda_{ii} \prod_{j>i} \extd \RE \lambda_{ij} \extd \IM\lambda_{ij}&= \prod_{i} \chi_{ii}^{-1} \prod_{j>i} (\chi_{ii}^2\chi_{jj}^2)^{-\frac{1}{2}}\prod_i \extd\sigma_{ii} \prod_{j>i} \extd \RE \sigma_{ij} \extd \IM\sigma_{ij}\\
& =(\det \, \chi)^{-\N} \, \prod_i \extd\sigma_{ii} \prod_{j>i} \extd \RE \sigma_{ij} \extd \IM\sigma_{ij}\,.
\end{align}
Furthermore:
\begin{equation}
(\det \lambda)^{-N}=(\det (\chi^{-1} \sigma))^{-N}=(\det \sigma)^{-N} (\det \chi)^{N}\,,
\end{equation}
and:
\begin{equation}
Z\propto \int \extd \chi \,e^{-\mu_1 \Tr\, \chi - \Tr\, V(\chi)}\,.
\end{equation}
In the large $\N$ limit, the integral is dominated by the saddle point. Using the polar decomposition of the integration measure over Hermitian matrices \cite{potters2020first,Francesco_1995}:
\begin{equation}
\extd \chi= \extd \Lambda (\Delta(\Lambda))^2 \extd U\,,
\end{equation}
where $\Lambda$ is a diagonal and positive definite matrix with eigenvalues $\lambda_i$: $\extd \Lambda \equiv \prod_i \extd \lambda_i$, $\Delta(\Lambda):=\prod_{j>i} (\lambda_j-\lambda_i)$ is the Vandermonde determinant and $\extd U$ is the Haar measure over the Unitary group. It is furthermore suitable to rescale $\chi \to \N \chi$, such that the saddle point equation reads:
\begin{equation}
\frac{1}{2}\left(\mu_1 + \sum_{p=1}^K \frac{p \mu_p}{(p!)^2}\, \lambda_i^{p-1}\right)=\frac{1}{N}\sum_{j\neq i}\, \frac{1}{\lambda_i-\lambda_j}\,.
\end{equation}
This equation can be investigated in the continuum limit from Tricomi's formula \cite{tricomi1985integral,potters2020first,landau1986theory}. Indeed, denoting as $\mu(\lambda)$ the large $\N$ eigenvalues distribution, the previous equation reads:
\begin{equation}
\underbrace{\frac{1}{2}\left(\mu_1 + \sum_{p=1}^K \frac{p \mu_p}{(p!)^2}\, \lambda^{p-1}\right)}_{f(\lambda)}=\dashint\, \extd\lambda^\prime\frac{\mu(\lambda^\prime)}{\lambda-\lambda^\prime}\,.
\end{equation}
If the spectrum is bounded, the solution for $\mu(\lambda)$ as the integral form:
\begin{equation}
\mu(\lambda):=-\frac{1}{\pi^2\sqrt{(\lambda-a)(b-\lambda)}}\left(\dashint\, \extd\lambda^\prime \sqrt{(\lambda^\prime-a)(b-\lambda^\prime)} \frac{f(\lambda^\prime)}{\lambda-\lambda^\prime}+C\right)\,,\label{formulamu}
\end{equation}
where $C$ is adjusted from the boundary and normalization conditions. Consider for instance a purely quadratic confining potential, for $f(\lambda)=\mu_1/2$, we have, choosing $C$ such that $\mu(b)=0$,
\begin{equation}
\mu(\lambda)=\frac{\mu_1\sqrt{\lambda (b-\lambda)}}{2 \pi \lambda}\,.
\end{equation}
Note that it is suitable to set $a=0$, and the boundary $b$ is then fixed by the normalization condition:
\begin{equation}
\int_0^b \extd \lambda \mu(\lambda)=1 \quad \to \quad b=\frac{4}{\mu_1}\,,
\end{equation}
and finally:
\begin{equation}
\mu(\lambda)=\frac{\mu_1 \sqrt{\lambda\left(\frac{4}{\mu_1}-\lambda\right)}}{2 \pi  \lambda}\,.
\end{equation}
Not surprisingly, the distribution looks like a Marchenko–Pastur distribution and the typical shape is shown in Figure \ref{FigMP}. 

The formula \eqref{formulamu} can be generalized in cases where we have only one hard wall or no hard wall. In the first case, assuming for instance we have a hard wall at $x=b$, the solution is \cite{landau1986theory}:
\begin{equation}
\mu(\lambda):=-\frac{1}{\pi^2} \sqrt{\frac{\lambda-a}{b-\lambda}}\dashint\, \extd\lambda^\prime \sqrt{\frac{b-\lambda^\prime}{\lambda^\prime-a}} \,\frac{f(\lambda^\prime)}{\lambda-\lambda^\prime}\,,
\end{equation}
and in the second case:
\begin{equation}
\mu(\lambda):=-\frac{1}{\pi^2} \sqrt{(\lambda-a)(b-\lambda)}\dashint\, \extd\lambda^\prime \frac{1}{\sqrt{(\lambda^\prime-a)(b-\lambda^\prime)}}\, \frac{f(\lambda^\prime)}{\lambda-\lambda^\prime}\,.
\end{equation}

\section{Fokker-Planck equation for Hermitian matrices}\label{App1}

In this appendix, we propose for the reader unfamiliar with the stochastic formalism to derive the Fokker-Planck equation for the Hermitian matrix (more detail about the general formalism can be found in the standard reference \cite{ZinnJustinBook2,Zinn-Justin:1989rgp}). Let us start with the definition of the probability $P[M,t]$:
\begin{equation}
P[M,t]:=\langle \delta \left(M-M(t)\right) \rangle\,,
\end{equation}
where the averaging is over the noise. It is easy to check that it is normalized to $1$ with respect to the integration measure over the Hermitian matrix space:
\begin{equation}
\int \extd M\,P[M,t]=\int \extd M \, \langle \delta \left(M-M(t)\right) \rangle =1\,,
\end{equation}
where $\extd M:= \prod_{i} \extd M_{ii}\prod_{j<i} \extd \RE M_{ij} \extd \IM M_{ij}$. We will prove that $P[M,t]$ satisfy to the equation:
\begin{equation}
\frac{d}{dt}P[M,t]=\textbf{H} P[M,t]\,. \label{}
\end{equation}

Let $\bf{q}\in \mathbb{R}^{\N^2}$ the real vector with $\N^2$ components, such that:
\begin{equation}
\textbf{q}= \{ M_{11},\cdots, M_{NN}, \RE M_{i>j},\cdots , \IM M_{i>j},\cdots  \}\,,
\end{equation}
the probability then can be rewritten in terms of the variable $\textbf{q}$ as $P(\textbf{q},t):=\langle \delta( \textbf{q}-\textbf{q}(t)) \rangle$. Now, because the system has no memory (the noise is randomly distributed with the same probability at each time), the probability as to be a Markovian process:
\begin{equation}
P[\textbf{q},t; \textbf{q}_0,t_0]=\int \, \extd^{N^2} \textbf{q}^\prime \, P[\textbf{q},t; \textbf{q}^\prime,t^\prime]P[\textbf{q}^\prime,t^\prime; \textbf{q}_0,t_0]\,,
\end{equation}
where in the notation $P[\textbf{q},t; \textbf{q}_0,t_0]$ we make appeared explicitly the initial condition $M(t_0)=M_0$. Because the origin of time is irrelevant, let us consider the probability $P[\textbf{q},t_0+\epsilon; \textbf{q}_0,t_0]$ to jump from $M_0$ to $M$ in the (expected small) time interval $\epsilon$. In the laps $\epsilon$, the trajectory changes as:
\begin{equation}
\textbf{q}(t_0+\epsilon)=-\textbf{f}(t_0) \epsilon +\int_{t_0}^{t_0+\epsilon}\extd t^\prime \,\textbf{b}(t^\prime)\,,
\end{equation}
where $\textbf{f}$ is the gradient force and $\textbf{b}$ is the vector in $\mathbb{R}^{\N^2}$ build from the noise matrix field $B$ as the vector $\textbf{q}$, explicitly for instance:
\begin{equation}
\textbf{f}=\left\{\frac{\delta \mathcal{H}}{\delta M_{11}}, \cdots , \frac{\delta \mathcal{H}}{\delta M_{NN}}, \RE \left(\frac{\delta \mathcal{H}}{\delta M_{1N}}\right), \cdots ,\IM\left(\frac{\delta \mathcal{H}}{\delta M_{1N}}\right), \cdots \right\}\,.\label{expfi}
\end{equation}
Now, we discretize the time interval into steps of size $\epsilon$, and we define the discrete noise $\textbf{b}_i$, for time $t_i=i \times \epsilon$ as:
\begin{equation}
\int_{t_0}^{t_0+\epsilon}\extd t^\prime \,\textbf{b}(t^\prime) = \sqrt{\epsilon}\, \textbf{b}_i\,,
\end{equation}
such that correlations between ${b}_{\alpha i}$ and ${b}_{\beta j}$ is:
\begin{equation}
\langle {b}_{\alpha i}{b}_{\beta j} \rangle = T\delta_{\alpha\beta}\delta_{ij}\,,
\end{equation}
where as before, indices $i,j \cdots$ refer to the time step, and Greeks indices $\alpha,\beta \cdots$ are the component of the vector $\textbf{b}_i$. The move equation from step $i$ to step $i+1$ then reads:
\begin{equation}
\textbf{q}_{i+1}=-\textbf{f}_i \epsilon +\sqrt{\epsilon} \textbf{b}_i\,. 
\end{equation}
The discrete probability $W(\epsilon,\textbf{q},\textbf{q}_0):= P[\textbf{q},t_0+\epsilon; \textbf{q}_0,t_0]$ to jump from $\textbf{q}_0$ to $\textbf{q}$ in the laps $\epsilon$ is:
\begin{align}
W(\epsilon,\textbf{q},\textbf{q}_0)&= \frac{1}{z_{\eta,i}}\int \extd^{\N^2} \textbf{b}_i\, e^{-\frac{\textbf{b}_{i\parallel}^2+2\textbf{b}_{i\bot}^2}{2T}} \delta (\textbf{q}-\textbf{q}_0+\textbf{f}_i \epsilon-\sqrt{\epsilon}\, \textbf{b}_i)\label{eqstepW}\\
&=\frac{1}{z_{\eta,i}} \exp{\left(-\frac{(\textbf{q}-\textbf{q}_0+\textbf{f}_i \epsilon)^2_{\parallel}}{2\epsilon T}-\frac{(\textbf{q}-\textbf{q}_0+\textbf{f}_i \epsilon)^2_\bot}{\epsilon T}\right)}\,,
\end{align}
where $\textbf{b}_{i\parallel}$ and $\textbf{b}_{i\bot}$, such that $\textbf{b}_{i\parallel}\oplus \textbf{b}_{i\bot}=\textbf{b}_{i}$, where:
\begin{align}
\textbf{b}_{i\parallel}&:=\{ M_{11},\cdots, M_{NN}\}\\
\textbf{b}_{i\bot}&:=\{\RE M_{i>j},\cdots , \IM M_{i>j},\cdots  \}\,.
\end{align}
Furthermore:
\begin{equation}
z_{\eta,i}:=(\epsilon)^{\frac{\N^2}{2}}\int \extd^{\N^2} \textbf{b}_i\, e^{-\frac{\textbf{b}_{i\parallel}^2+2\textbf{b}_{i\bot}^2}{2T}} \,.
\end{equation}
Now, taking the Fourier transform with respect to the variable $\textbf{q}$, we get:
\begin{align}
&e^{-\frac{\epsilon T}{2}\textbf{p}^2_{\parallel}+i\textbf{f}_{i\parallel}\cdot \textbf{p}_{\parallel} \epsilon-i \textbf{q}_{0\parallel}\cdot \textbf{p}_{\parallel}}=e^{-i\textbf{q}_{0\parallel}\cdot \textbf{p}_{\parallel}} \left(1-\frac{\epsilon T}{2} \textbf{p}^2_{\parallel}+i \textbf{f}_{i\parallel}\cdot \textbf{p}_{\parallel} \epsilon\right) +\mathcal{O}(\epsilon^2)\,\\
&e^{-\frac{\epsilon T}{4}\textbf{p}^2_{\bot}+i\textbf{f}_{i\bot}\cdot \textbf{p}_{\parallel} \epsilon-i \textbf{q}_{0\bot}\cdot \textbf{p}_{\bot}}=e^{-i\textbf{q}_{0\bot}\cdot \textbf{p}_{\bot}} \left(1-\frac{\epsilon T}{4} \textbf{p}^2_{\bot}+i \textbf{f}_{i\bot}\cdot \textbf{p}_{\bot} \epsilon\right) +\mathcal{O}(\epsilon^2)\,.
\end{align}
The linear term in $\epsilon$ is nothing but the matrix element $\langle \textbf{p} \vert \textbf{H} \vert \textbf{q}_0 \rangle$, then, taking the inverse Fourier transform, we get the matrix element $\langle \textbf{q} \vert \textbf{H} \vert \textbf{q}_0 \rangle$,
\begin{equation}
\langle \textbf{q} \vert \textbf{H} \vert \textbf{q}_0 \rangle=\left(\frac{T}{2}\mathcal{D}+ \sum_\alpha\frac{\partial}{\partial q_\alpha}f_{\alpha i} \right)\delta^{\N^2}(\textbf{q}-\textbf{q}_0)\,,
\end{equation}
where:
\begin{align}
\mathcal{D} &:=\frac{\partial^2}{\partial \textbf{q}_{i \parallel}^2}+\frac{1}{2}\frac{\partial^2}{\partial \textbf{q}_{i \bot}^2}\\
&:= \sum_{i} \frac{\partial^2}{\partial M_{ii}^2}+\frac{1}{2}\sum_{i<j}\left(\frac{\partial^2}{\partial (\RE(M)_{ij})^2}+\frac{\partial^2}{\partial (\IM(M)_{ij})^2}\right)\,,
\end{align}
is the unitary invariant Laplacian operator on Hermitian matrices \cite{itzykson1980planar}. We then proved that:
\begin{equation}
\frac{\partial P[\textbf{q},t]}{\partial t}=\textbf{H} P[\textbf{q},t]\,,
\end{equation}
with, returning to the matrix notation and using the explicit expression for the driving force \eqref{expfi}, we get:
\begin{equation}
\textbf{H}:=\frac{1}{2}\sum_{i,j}\left(T \frac{\partial }{\partial{M_{ij}}\partial {\overline{M}_{ij}}}+2\frac{\partial^2 H }{\partial{M_{ij}}\partial {\overline{M}_{ij}}} + \frac{\partial H }{\partial{M_{ij}}}\frac{\partial}{\partial {\overline{M}_{ij}}}+\frac{\partial H }{\partial{\overline{M}_{ij}}}\frac{\partial}{\partial {{M}_{ij}}} \right)\,,
\end{equation}
where we make use of the identity
\begin{equation}
4\frac{\partial}{\partial z}\frac{\partial}{\partial{\bar{z}}}\equiv \frac{\partial^2}{\partial x^2}+\frac{\partial^2}{\partial y^2}\,,\label{relationcomplexderiv}
\end{equation}
and used the definition of $H$ (see equation \eqref{HamiltonianCdef}, without index $\infty$ because large $\N$ limit is not required):
\begin{equation}
\frac{\delta \mathcal{H}}{\delta M_{ij}(t)}\equiv \frac{\partial H}{\partial M_{ij}(t)}\,.
\end{equation}
Finally, we easily check that $e^{-2\mathcal{H}/T}$ is an eigenvector with vanishing eigenvalue:
\begin{equation}
 \textbf{H} \exp \left(-2\frac{H[M]}{T}\right)=0\,.
\end{equation}

\printbibliography[heading=bibintoc]

@article{zinn1998vector,
  title={Vector models in the large $ N $ limit: a few applications},
  author={Zinn-Justin, Jean},
  journal={arXiv preprint},
doi="https://doi.org/10.48550/arXiv.hep-th/9810198",
  year={1998}
}

@article{blaizot2006nonperturbative,
  title={Nonperturbative renormalization group and momentum dependence of n-point functions. II},
  author={Blaizot, Jean-Paul and Mendez-Galain, Ramon and Wschebor, Nicolas},
  journal={Physical Review E},
doi="10.1103/PhysRevE.74.051116",
  volume={74},
  number={5},
  pages={051117},
  year={2006},
  publisher={APS}
}

@article{kpera2023anomalous,
  title={Anomalous higher order Ward identities in tensorial group field theories without closure constraint},
  author={Kpera, Bio Wahabou and Lahoche, Vincent and Ousmane Samary, Dine  and Yerima, Seke Fawaaz Zime},
  journal={arXiv preprint},
doi="10.48550/arXiv.2307.12446",
  year={2023}
}

@article{lahoche20241,
  title={Functional renormalization group for “p = 2” like
glassy matrices in the planar approximation: 1 Vertex expansion at equilibrium},
  author={Lahoche, Vincent and Ousmane Samary, Dine},
  journal={arXiv preprint},
doi="arXiv:2403.07577",
  year={2024}
}

@article{otto2022regulator,
  title={Regulator scheme dependence of the chiral phase transition at high densities},
  author={Otto, Konstantin and Busch, Christopher and Schaefer, Bernd-Jochen},
  journal={Physical Review D},
doi="10.1103/PhysRevD.106.094018",
  volume={106},
  number={9},
  pages={094018},
  year={2022},
  publisher={APS}
}

@article{lahoche2016renormalization,
    author = "Lahoche, Vincent and Oriti, Daniele",
    title = "{Renormalization of a tensorial field theory on the homogeneous space SU(2)/U(1)}",
    eprint = "1506.08393",
    archivePrefix = "arXiv",
    primaryClass = "hep-th",
    doi = "10.1088/1751-8113/50/2/025201",
    journal = "J. Phys. A",
    volume = "50",
    number = "2",
    pages = "025201",
    year = "2017"
}

@article{samary2014closed,
    author = "Ousmane Samary, Dine",
    title = "{Closed equations of the two-point functions for tensorial group field theory}",
    eprint = "1401.2096",
    archivePrefix = "arXiv",
    primaryClass = "hep-th",
    doi = "10.1088/0264-9381/31/18/185005",
    journal = "Class. Quant. Grav.",
    volume = "31",
    pages = "185005",
    year = "2014"
}

@book{forrester2010log,
  title={Log-gases and random matrices (LMS-34)},
  author={Forrester, Peter J},
  year={2010},
  publisher={Princeton University Press}
}

@article{Lahoche:2018ggd,
    author = "Lahoche, Vincent and Ousmane Samary, Dine",
    title = "{Ward identity violation for melonic $T^4$-truncation}",
    eprint = "1809.06081",
    archivePrefix = "arXiv",
    primaryClass = "hep-th",
    doi = "10.1016/j.nuclphysb.2019.01.005",
    journal = "Nucl. Phys. B",
    volume = "940",
    pages = "190--213",
    year = "2019"
}

@article{Lahoche:2018oeo,
    author = "Lahoche, Vincent and Ousmane Samary, Dine",
    title = "{Nonperturbative renormalization group beyond melonic sector: The Effective Vertex Expansion method for group fields theories}",
    eprint = "1809.00247",
    archivePrefix = "arXiv",
    primaryClass = "hep-th",
    doi = "10.1103/PhysRevD.98.126010",
    journal = "Phys. Rev. D",
    volume = "98",
    number = "12",
    pages = "126010",
    year = "2018"
}

@book{Zinn-Justin:1989rgp,
    author = "Zinn-Justin, Jean",
    title = "{Quantum field theory and critical phenomena}",
    isbn = "978-0-19-850923-3, 978-0-19-883462-5",
    publisher = "Oxford University Press",
    series = "International Series of Monographs on Physics",
doi="10.1093/acprof:oso/9780198509233.001.0001",
    volume = "77",
    month = "4",
    year = "2021"
}

@book{ZinnJustinBook2,
        author = {J. Zinn-Justin
},
        title = {From random walks to random matrices},
        publisher= {Oxford Graduate Texts},
doi="10.1093/oso/9780198787754.001.0001",
        year = {2019}
}

@book{landau1986theory,
  title={Theory of elasticity: volume 7},
  author={Landau, Lev Davidovich and Lifshitz, Evgenii Mikhailovich and Kosevich, Arnolʹd Markovich and Pitaevskii, Lev Petrovich},
doi="10.1016/C2009-0-25521-8",
  volume={7},
  year={1986},
  publisher={Elsevier}
}

@book{tricomi1985integral,
  title={Integral equations},
  author={Tricomi, Francesco Giacomo},
  volume={5},
  year={1985},
  publisher={Courier corporation}
}

@article{Francesco_1995,
	doi = {10.1016/0370-1573(94)00084-g},
  
	url = {https://doi.org/10.1016%2F0370-1573%2894%2900084-g},
  
	year = 1995,
	month = {mar},
  
	publisher = {Elsevier {BV}},
  
	volume = {254},
  
	number = {1-2},
  
	pages = {1--133},
  
	author = {P.Di Francesco and P. Ginsparg and J. Zinn-Justin},
  
	title = {2D gravity and random matrices},
  
	journal = {Physics Reports}
}

@book{guruau2017random,
  title={Random tensors},
  author={Gur{\u{a}}u, R{\u{a}}zvan Gheorghe},
  year={2017},
  publisher={Oxford University Press},
doi="10.1093/acprof:oso/9780198787938.001.0001"
}

@article{lahoche2022generalized,
  title={Generalized scale behavior and renormalization group for data analysis},
  author={Lahoche, Vincent and Samary, Dine Ousmane and Tamaazousti, Mohamed},
  journal={Journal of Statistical Mechanics: Theory and Experiment},
doi="10.1088/1742-5468/ac52a6",
  volume={2022},
  number={3},
  pages={033101},
  year={2022},
  publisher={IOP Publishing}
}

@article{lahoche2023functional,
  title={Functional renormalization group for multilinear disordered Langevin dynamics II: Revisiting the p= 2 spin dynamics for Wigner and Wishart ensembles},
  author={Lahoche, Vincent and Ousmane Samary, Dine and Tamaazousti, Mohamed},
  journal={J. Phys. Comm},
doi="10.1088/2399-6528/acd09d",
  year={2023}
}

@article{Lahoche_2020b,
    author = "Lahoche, Vincent and Ousmane Samary, Dine",
    title = "{Pedagogical comments about nonperturbative Ward-constrained melonic renormalization group flow}",
    eprint = "2001.00934",
    archivePrefix = "arXiv",
    primaryClass = "hep-th",
    doi = "10.1103/PhysRevD.101.024001",
    journal = "Phys. Rev. D",
    volume = "101",
    number = "2",
    pages = "024001",
    year = "2020"
}

@article{Lahoche_2019bb,
    author = "Lahoche, Vincent and Ousmane Samary, Dine",
    title = "{Ward identity violation for melonic $T^4$-truncation}",
    eprint = "1809.06081",
    archivePrefix = "arXiv",
    primaryClass = "hep-th",
    doi = "10.1016/j.nuclphysb.2019.01.005",
    journal = "Nucl. Phys. B",
    volume = "940",
    pages = "190--213",
    year = "2019"
}

@article{Carrozza_2017,
	doi = {10.1103/physrevd.96.066007},
  
	url = {https://doi.org/10.1103%2Fphysrevd.96.066007},
  
	year = 2017,
	month = {sep},
  
	publisher = {American Physical Society ({APS})},
  
	volume = {96},
  
	number = {6},
  
	author = {Sylvain Carrozza and Vincent Lahoche and Daniele Oriti},
  
	title = {Renormalizable group field theory beyond melonic diagrams: An example in rank four},
  
	journal = {Physical Review D}
}

@article{Carrozza_2017a,
	doi = {10.1088/1361-6382/aa6d90},
  
	url = {https://doi.org/10.1088%2F1361-6382%2Faa6d90},
  
	year = 2017,
	month = {may},
  
	publisher = {{IOP} Publishing},
  
	volume = {34},
  
	number = {11},
  
	pages = {115004},
  
	author = {Sylvain Carrozza and Vincent Lahoche},
  
	title = {Asymptotic safety in three-dimensional {SU}(2) group field theory: evidence in the local potential approximation},
  
	journal = {Classical and Quantum Gravity}
}

@article{Carrozza_2016ccc,
	doi = {10.3842/sigma.2016.070},
  
	url = {https://doi.org/10.3842%2Fsigma.2016.070},
  
	year = 2016,
	month = {jul},
  
	publisher = {{SIGMA} (Symmetry, Integrability and Geometry: Methods and Application)},
  
	author = {Sylvain Carrozza},
  
	title = {Flowing in Group Field Theory Space: a Review},
  
	journal = {Symmetry, Integrability and Geometry: Methods and Applications}
}

@incollection{Delamotte_2012,
	doi = {10.1007/978-3-642-27320-9_2},
  
	url = {https://doi.org/10.1007%2F978-3-642-27320-9_2},
  
	year = 2012,
	publisher = {Springer Berlin Heidelberg},
  
	pages = {49--132},
  
	author = {Bertrand Delamotte},
  
	title = {An Introduction to the Nonperturbative Renormalization Group},
  
	booktitle = {Renormalization Group and Effective Field Theory Approaches to Many-Body Systems}
}

@article{Berges_2002,
	doi = {10.1016/s0370-1573(01)00098-9},
  
	url = {https://doi.org/10.1016%2Fs0370-1573%2801%2900098-9},
  
	year = 2002,
	month = {jun},
  
	publisher = {Elsevier {BV}},
  
	volume = {363},
  
	number = {4-6},
  
	pages = {223--386},
  
	author = {Jürgen Berges and Nikolaos Tetradis and Christof Wetterich},
  
	title = {Non-perturbative renormalization flow in quantum field theory and statistical physics},
  
	journal = {Physics Reports}
}

@article{benedetti2016functional,
  title={Functional renormalization group approach for tensorial group field theory: a rank-6 model with closure constraint},
  author={Benedetti, Dario and Lahoche, Vincent},
  journal={Classical and Quantum Gravity},
doi="10.1088/0264-9381/33/9/095003",
  volume={33},
  number={9},
  pages={095003},
  year={2016},
  publisher={IOP Publishing}
}

@article{canet2003nonperturbative,
  title={Nonperturbative renormalization group approach to the Ising model: a derivative expansion at order $\partial^4$},
  author={Canet, L{\'e}onie and Delamotte, Bertrand and Mouhanna, Dominique and Vidal, Julien},
  journal={Physical Review B},
doi="10.1103/PhysRevB.68.064421",
  volume={68},
  number={6},
  pages={064421},
  year={2003},
  publisher={APS}
}

@book{potters2020first,
  title={A First Course in Random Matrix Theory: For Physicists, Engineers and Data Scientists},
  author={Potters, Marc and Bouchaud, Jean-Philippe},
  year={2020},
  publisher={Cambridge University Press},
doi="10.1017/9781108768900"
}

@article{itzykson1980planar,
  title={The planar approximation. II},
  author={Itzykson, Claude and Zuber, J-B},
  journal={Journal of Mathematical Physics},
doi="10.1063/1.524438
",
  volume={21},
  number={3},
  pages={411--421},
  year={1980},
  publisher={American Institute of Physics}
}
\end{document}